%% file: main.tex
\documentclass[%
aip,
amsmath,amssymb,
reprint,%
]{revtex4-1}
\usepackage{graphicx}% Include figure files
\usepackage{dcolumn}% Align table columns on decimal point
\usepackage{bm}% bold math
\usepackage[utf8]{inputenc}
\usepackage[T1]{fontenc}
\usepackage{mathptmx}
\usepackage{soul}
\usepackage{etoolbox}
\usepackage{multirow}
\usepackage{xspace}
\usepackage{tablefootnote}
\usepackage{soul}
\makeatletter
\def\@email#1#2{%
 \endgroup
 \patchcmd{\titleblock@produce}
  {\frontmatter@RRAPformat}
  {\frontmatter@RRAPformat{\produce@RRAP{*#1\href{mailto:#2}{#2}}}\frontmatter@RRAPformat}
  {}{}
}%

\usepackage[mathscr]{euscript}
\usepackage{color}
\usepackage{xcolor}
\usepackage{xspace}
\usepackage{amsmath}

\makeatother
\begin{document}
\input{symbols/main}

%------------------
\preprint{APS/123-QED}

\title{
Cross-correlation effects in the solution NMR spectra %spectra 
of near-equivalent
%$\mathrm{{}^{13}C}$ 
spin-1/2 pairs
%I  Spectral lineshapes.
}
\author{James W. Whipham}
\author{Gamal Moustafa}
\author{Mohamed Sabba}
\author{Weidong Gong}
\author{Christian Bengs}
\author{Malcolm H. Levitt}
 \email{mhl@soton.ac.uk}

\affiliation{Department of Chemistry, University of Southampton, SO17 1BJ, UK}%

\date{\today}% 

\begin{abstract}
The NMR spectra of spin-1/2 pairs contains four peaks, with two inner peaks much stronger than the outer peaks in the near-equivalence regime. We have observed that the strong inner peaks have significantly different linewidths, when measurements were performed on a \Ctwo-labelled triyne derivative. This linewidth difference may be attributed to strong cross-correlation effects. We develop the theory of cross-correlated relaxation in the case of near-equivalent homonuclear spin-1/2 pairs, in the case of a molecule exhibiting strongly anisotropic rotational diffusion. Good agreement is found with the experimental NMR lineshapes.
\end{abstract}

\maketitle

%tableofcontents

%\MHLnote{Mohamed, Gamal and Weidong should be included in author list. Possibly Christian as well. Discuss.}

%---------------------------------------
\section{Introduction}
%-------------
If a nuclear spin system is perturbed from a thermal equilibrium state, it slowly returns to equilibrium through nuclear spin relaxation. Such relaxation processes are driven by fluctuations in the interactions between the nuclear spins and the thermal molecular environment. In general, many types of fluctuating interaction are involved, and these interactions may be correlated with each other. For example, in solution NMR, the fluctuations of nuclear spin interactions are caused by random molecular tumbling, and since the rotation of a molecule modulates all intramolecular interactions at the same time, the fluctuations of these interactions are correlated. Such cross-correlation effects are well-documented in solution NMR~\cite{mcconnell_effect_1956, shimizu_theory_1964,werbelow_intramolecular_1977,goldman_interference_1984,di_bari_magnetization_1990,kumar_cross-correlations_2000,madhu_cross-correlation_2002,kowalewski_book_2018}. Cross-correlation gives rise to differential line broadening and line narrowing, and differences in the longitudinal relaxation behaviour of individual multiplet components~\cite{mcconnell_effect_1956,shimizu_theory_1964,werbelow_intramolecular_1977,goldman_interference_1984,kumar_cross-correlations_2000,kowalewski_book_2018}. Cross-correlation effects have been used to estimate the relative orientations of nuclear spin interaction tensors, allowing the estimation of molecular torsional angles~\cite{reif_direct_1997,reif_determination_1998,ravindranathan_investigation_2000}. A particularly important set of cross-correlation effects is associated with the so-called TROSY techniques (transverse relaxation-optimized spectroscopy), which have important applications, especially in biomolecular NMR~\cite{pervushin_TROSY_1997,tugarinov_MeTROSY_2003}.

Cross-correlation often takes place between the fluctuations of internuclear dipole-dipole (DD) couplings and chemical shift anisotropy (CSA) interactions. Such DD-CSA cross-correlation effects are well-known for heteronuclear spin pairs, and underpin important techniques such as TROSY~\cite{goldman_interference_1984,pervushin_TROSY_1997,tugarinov_MeTROSY_2003}. In this paper we demonstrate strong DD-CSA cross-correlation effects in the solution NMR of a system containing \emph{homonuclear} pairs of \Cth nuclei, in the limit of \emph{near-magnetic-equivalence}, implying that the chemical shift difference between the coupled nuclear sites is much smaller than the internuclear J-coupling.

The system of interest is the \Ctwo-labelled triyne derivative referred to here as \One, which has the following systematic name: 1-methoxy-4-((4-methoxymethoxy)phenyl)hexa-1,3,5-triyn-1-yl)benzene. Its molecular structure is shown in figure~\ref{fig:triyne_and_tensors}a. Each molecule of \One has a rod-like shape, with two \Cth labels at the central pair of carbon atoms, in the centre of the triyne moiety. The two end groups are different, endowing the two \Cth nuclei with slightly different chemical shifts ($\Delta\delta=0.16$ ppm). Since the \Cth-\Cth J-coupling is large ($\JCC=214.15$ Hz), the \Cth pair is in the near-equivalent regime at all accessible magnetic fields~\cite{tayler_singlet_2011}.

The \Cth NMR spectrum of a 0.3 M solution of \One in \(\mathrm{CDCl}_3\) is shown in figure~\ref{fig:full_and_superimposed_spectrum}. This corresponds to the expected AB four-peak structure, although the two outer peaks are too weak to be observed. The two strong central peaks are only partially resolved, and form a strongly asymmetric lineshape, as shown by the inset in figure~\ref{fig:full_and_superimposed_spectrum}. As discussed below, the asymmetry of the central peak pair is due to strong DD-CSA cross-correlation effects. 

An analysis of cross-correlated relaxation in \One must take into account its rod-like shape, which causes strongly anisotropic rotational diffusion in solution. The theory of nuclear spin relaxation has been developed in the context of model-free treatments of biomolecules with anisotropic internal motions~\cite{lipari_model-free_1982,lipari_model-free_1982-1,lee_rotational_1997,marcellini_accurate_2020,tjandra_anisotropic_1996,osborne_anisotropic_2001}. However, most existing treatments of cross-correlated relaxation in small molecules assume approximately isotropic rotational diffusion, which is clearly not applicable here. In the following sections we develop the theory of cross-correlated relaxation in systems with anisotropic rotational diffusion. We provide analytical formulae for the NMR spectrum of a near-equivalent homonuclear spin pair undergoing cross-correlated relaxation in the presence of anisotropic rotational diffusion. The observed spectral asymmetry is reproduced, with good agreement between theory, experiment and numerical simulations. 

%This paper concerns the asymmetric NMR spectrum of \One in solution. In the following paper, we examine the transition-dependent longitudinal relaxation of \One, which is also strongly affected by cross-correlated relaxation. 
%\MHLnote{The comment about the next paper has been removed and needs to be inserted into the discussion at the end.}

\input{figures/triyne_and_tensors}

%==============================

\section{Experimental}

\input{figures/full_and_superimposed_spectrum}

\subsection{Sample}

The synthesis of \One is described in the supplementary material. 19 mg of \One was made up to a 200 \(\mu\mathrm{L}\) 0.3 M solution in \CDCl3. 5 freeze-thaw degassing cycles were performed on the solution.
\subsection{NMR}

The experiments were performed on a 400 MHz (9.4 T) Bruker Avance Neo spectrometer. The pulse sequence was a simple 90\textdegree \  pulse-acquire. The \(^{13}\mathrm{C}\) nutation  frequency was 6.68 kHz and 1 scan was performed. The NMR signal was sampled with 131 k data points with a spectral width of 81.46 ppm.

\subsection{Computational chemistry}

Geometry optimisation and simulation of the magnetic shielding tensors of \One were performed at the B3LYP/aug-cc-PVTZ~\cite{becke_densityfunctional_1992,dunning_gaussian_1989,kendall_electron_1992} level of theory in the Gaussian 09 suite of programs~\cite{g09}. After geometry optimisation, the dipole-dipole coupling tensor between the two \Cth nuclei was calculated from the internuclear distance. The parameters obtained from the computations are presented in table~\ref{tab:parameters}.

\input{tables/parameters}
\section{Theory}

\subsection{Anisotropic rotational diffusion}

To analyze the relaxation behaviour of this system, \One is approximated as a rigid molecule undergoing anisotropic rotational diffusion in solution, and is presented in the vein of Huntress~\cite{huntress_effects_1968,huntress_study_1970}. We treat the molecule as a rod-shaped symmetric top, corresponding to a strongly anisotropic inertia tensor depicted by the ovaloid~\cite{radeglia_ovaloid_1995,young_tensorview_2019} shown in figure~\ref{fig:triyne_and_tensors}d, with a rotational diffusion tensor coincident with the inertia tensor.

The ovaloid representation of the inertia tensor, shown in figure~\ref{fig:triyne_and_tensors}d, has the form of a dimpled disk, or a doughnut with an incomplete hole. This shape may be interpreted as follows: Take a vector starting from the molecular centre of mass, and pointing in any direction. The vector intersects the ovaloid surface at some point. The distance from the centre of mass to the intersection point is proportional to the moment of inertia for a rotation around that vector. Rotations around an axis which is perpendicular to the long axis of the molecule are associated with a large moment of inertia, so that the surface is relatively distant from the centre of mass. A rotation around the long axis, on the other hand, has a small moment of inertia, so that the surface is close to the origin in that direction. Hence the ovaloid has the appearance of a dimpled disk, with the dimples along the long axis of the molecule.

The principal axis system of the rotational diffusion tensor is denoted $D$ and is depicted in figure~\ref{fig:frame_transformations}. A laboratory reference frame $L$ may be defined such that its z-axis is aligned with the external magnetic field (see figure~\ref{fig:frame_transformations}). The relative orientation of the $D$ and $L$ frames may be described by using the Euler angle triplet $\WDL=\{\aDL,\bDL,\gDL\}$. In this article, the z-y-z convention for Euler angles is used throughout~\cite{millot_active_2012}. Due to the molecular tumbling, these Euler angles are in general stochastic functions of time, $\WDL=\WDL(t)$, since the $D$-frame is molecule-fixed whilst the $L$-frame is space-fixed.

The stochastic time-dependence of the Euler angles $\WDL$ may be expressed in terms of the time-correlation functions of the rank-$l$ Wigner D-matrices:
\begin{equation} \label{time_corr_function}
\begin{split}
    G^{ll'}_{mm'nn'}(t_0,\tau) = & \overline{D^{(l)}_{mn}(\WDL(t_0)) D^{(l')*}_{m'n'}( \WDL(t_0+\tau))} 
\\
    = & \int \int d\WDL(t_0){\ } d\WDL(t_0+\tau)
\\
   & \times D^{(l)}_{mn}(\WDL(t_0)) D^{(l')*}_{m'n'}(\WDL(t_0+\tau)) \\
   & \quad \quad \times
P(\WDL(t_0)) P(\WDL(t_0+\tau)|\WDL(t_0))
\end{split}
\end{equation}
\noindent where \(P(\WDL(t_0)) = (8\pi^{2})^{-1}\) is the probability density that the molecule hosting the spin system will be at orientation \(\WDL(t_0)\) at time \(t=t_0\), and \(P(\WDL(t_0+\tau)|\WDL(t_0))\) is the conditional probability that the molecule will be at orientation \(\WDL(t_0+\tau)\) at time \(t = t_0+\tau\), given that it was at orientation \(\WDL(t_0)\) at time \(t=t_0\). If the stochastic process is assumed to be stationary, these probabilities are independent of $t_0$, allowing the arbitrary choice of time origin $t_0=0$. The expression for the conditional probability is given by Favro~\cite{favro_theory_1960}, and is,
\begin{equation} \label{conditional_probability}
    P(\Omega_{DL}(\tau)|\Omega_{DL}(0)) = \sum_{\nu} \psi^{*}_{\nu}(0) \psi_{\nu}(\tau) e^{-E_{\nu}\tau},
\end{equation}
where \(\psi_{\nu}(t)\) are eigenfunctions of the operator \(H_{\mathrm{rot\,diff}} = \mathbf{L} \cdot \mathbf{D} \cdot \mathbf{L}\), with corresponding eigenvalues \(E_{\nu}\), where \(\mathbf{L}\) and \(\mathbf{D}\) are the angular momentum vector and the diffusion tensor, respectively. For a symmetric top, we may write,
\begin{equation} \label{H_rot_diff}
    H_{\mathrm{rot\,diff}} = D_{\perp} \left( L^{2}_{x} + L^{2}_{y} \right) + D_{\parallel}L^{2}_{z},
\end{equation}
where \(D_{\perp}\) and \(D_{\parallel}\) are rotational diffusion constants associated with axes perpendicular and parallel, respectively, with the molecular long axis. Note that eq. (\ref{H_rot_diff}) is written in the \(D\)-frame.

Eq. (\ref{H_rot_diff}) is of the same form as the rigid-rotor Hamiltonian for a symmetric-top. As such, the eigenfunctions and eigenvalues in eq. (\ref{conditional_probability}) are those of a quantum mechanical rigid-rotor~\cite{zare_angular_1988,wang_anisotropic-rotational_1973}:
\begin{equation}
    \psi_{\nu} (t) \xrightarrow[]{} \phi^{J}_{K,M}(t) \equiv (-1)^{M-K}\sqrt{\frac{2J+1}{8\pi^{2}}} D^{(J)}_{-M-K}(\Omega(t))
\end{equation}
\begin{equation}
    E_{\nu} \xrightarrow[]{} E^{J}_{K} \equiv D_{\perp} J( J + 1 ) + K^{2} ( D_{\parallel} - D_{\perp} ).
\end{equation}
For our specific system, we will see that \(J \equiv l = 2\) and \(K = 0\). The only non-vanishing term in the correlation function is then \(E^{(2)}_{0} = 6 D_{\perp}\), and related to the rotational correlation time \(\tau_{\perp} \equiv \big(6 D_{\perp} \big)^{-1}\). Rotational motion around the molecular long axis does not modulate the interactions responsible for relaxation. This is a consequence of the coincidence of the \(P\)- and \(D\)-frames. The secularised time-correlation function 
%for our system 
becomes,
\begin{equation} \label{time_corr_function}
    G^{l}_{mnn'}(\tau) =
    G^{ll'}_{mm'nn'}(0,\tau)= \delta_{ll'}\delta_{mm'}\frac{(-1)^{n+n'}}{2l+1} e^{-\tau/\tau_{\perp}}.
\end{equation}
%
%On the other hand, Euler angles describing the relative orientation of the $P$-frame of a given interaction and the $D$-frame are time-independent, since both are molecule-fixed frames which tumble with the molecule. It follows that any transformation between a $P$- and the $D$-frame are also stochastic functions of time.

%These frame transformations are used in section \ref{fluctuating_hamiltonian_section} to express the spatial and spin part of the spin Hamiltonian in the laboratory frame, and section~\ref{relax_superop_section} to simplify the relaxation superoperator.

\input{figures/frame_transformations}

\subsection{Coherent Hamiltonian} \label{coherent_H_section}

The coherent spin Hamiltonian describes those spin interactions which are the same for all members of the spin ensemble at a given point in time. For a homonuclear spin-1/2 pair in solution, it may be written in the rotating frame of the Zeeman interaction as,
\begin{equation} \label{coh_ham}
    H_{\mathrm{coh}} = \frac{1}{2}\omega_{\Sigma}(I_{1z} + I_{2z}) + \frac{1}{2}\omega_{\Delta}(I_{1z} - I_{2z}) + \omega_{J}\mathbf{I}_{1} \cdot \mathbf{I}_{2},    
\end{equation}
\noindent with%,
\begin{align}
    \omega_{\Sigma} & = \omega_{1} + \omega_{2} 
\nonumber\\
    \omega_{\Delta} & = \omega_{1} - \omega_{2} 
\nonumber\\
    \omega_{J} & = 2 \pi J_{12},
\end{align}
\noindent where \(J_{12}\) is the isotropic part of the spin-spin coupling tensor, and \(\omega_{j}\) is the precession frequency of spin $I_j$,
%
%the \(j^{\mathrm{th}}\) spin \st{given by},
%
\begin{equation}
    \omega_{j} = \omega_{0}(1 + \delta_{j}^\mathrm{iso}) - \omega_{\mathrm{rf}}.
\end{equation}

\noindent Here, \(\omega_{0}\) is the Larmor frequency of the isotope, \(\delta_{j}^\mathrm{iso}\) is the isotropic chemical shift for the \(j^{\mathrm{th}}\) spin, and \(\omega_{\mathrm{rf}}\) is the radiofrequency carrier frequency.

The Hamiltonian may be diagonalised by using the perturbed singlet-triplet basis, $\BSTp$, defined as,
\begin{equation}\label{eq:BSTp}
\BSTp = \left\{ | S_{0}'\rangle, |T_{+1}'\rangle, |T_{0}'\rangle, |T_{-1}'\rangle
\right\},
\end{equation}
with elements,
\begin{align}
    & |S_{0}'\rangle = \mathrm{cos}\frac{\theta}{2}|S_{0}\rangle - \mathrm{sin}\frac{\theta}{2}|T_{0}\rangle 
\nonumber\\
    & |T_{+1}'\rangle = |T_{+1}\rangle 
\nonumber\\
    & |T_{0}'\rangle = \mathrm{sin}\frac{\theta}{2}|S_{0}\rangle + \mathrm{cos}\frac{\theta}{2}|T_{0}\rangle 
\nonumber\\
    & |T_{-1}'\rangle = |T_{-1}\rangle,
\end{align}
where the singlet and triplet states are given by,
\begin{align}
    & | S_{0} \rangle = \frac{1}{\sqrt{2}} \big( | \alpha \beta \rangle - | \beta\alpha \rangle \big) 
\nonumber\\
    & | T_{+1} \rangle = | \alpha\alpha \rangle
\nonumber\\
    & | T_{0} \rangle = \frac{1}{\sqrt{2}} \big( | \alpha\beta \rangle + | \beta\alpha \rangle \big) 
\nonumber\\
    & | T_{-1} \rangle = | \beta\beta \rangle,
\end{align}
and \(\theta\) is the \textit{singlet-triplet mixing angle}, defined by,
\begin{equation}
\mathrm{tan}\theta = {\omega_{\Delta}}/\omega_{J}.    
\end{equation}
The eigenvalues of \(H_{\mathrm{coh}}\) are, %\st{then},
\begin{align}
   % E_{S_{0}'}
   \omega_{S'_{0}}
%\omega\!\left(S_{0}'\right)
    & = -\frac{1}{4}\big( \omega_{J} + 2 \omega_{e} \big)
\nonumber\\
\omega_{T'_{+1}}%\omega\!\left(T_{+1}'\right)
%    E_{T_{+1}'} 
    & = +\frac{1}{4}\big( \omega_{J} + 2 \omega_{\Sigma} \big) 
\nonumber\\
\omega_{T'_{0}}%\omega\!\left(T_{0}'\right)
%    E_{T_{0}'} 
    & = -\frac{1}{4}\big( \omega_{J} - 2 \omega_{e} \big) 
\nonumber\\
\omega_{T'_{-1}}%\omega\!\left(T_{-1}'\right)
%   E_{T_{-1}'} 
    & = +\frac{1}{4} \big( \omega_{J} - 2 \omega_{\Sigma} \big),
\end{align}
where,
\begin{equation}\label{eq:we}
    \omega_e^2 = 
    \omega^{2}_{\Delta} + \omega^{2}_{J}.
\end{equation}

These eigenvalues are used in section \ref{liouvillian_section} to analyse the signal, allowing assignment of coherence-peak correspondence.

%======================
%\subsubsection{Fluctuating Hamiltonian}
\subsection{Fluctuating Hamiltonian} \label{fluctuating_hamiltonian_section}

%The fluctuating Hamiltonian is a sum of those Hamiltonians for interactions which differ between identical members of the spin ensemble.\CBnote{
The fluctuating Hamiltonian is a sum of contributions from the anisotropic spin interactions. These interactions differ between ensemble members at a given point in time, due to the random molecular tumbling. 
%In this paper, this \CBnote{our analysis?} 
The current analysis is restricted to the intra-pair dipole-dipole (DD) and chemical shift anisotropy (CSA) interactions,
\begin{equation} \label{fluctuating_hamiltonian}
    H_{\mathrm{fluc}} = H_{\mathrm{DD}} + H_{\mathrm{CSA}},
\end{equation}
as well as the cross-correlation between the two mechanisms.

The spin Hamiltonian for interaction \(\Lambda\) may be written in terms of irreducible spherical tensor operators as~\cite{smith_hamiltonians_1992},
\begin{equation} \label{H_tensors}
    H_{\Lambda}(t) = 
    c_{\Lambda} 
    \sum_{l=0}^{+2} \ %
    \sum_{m=-l}^{+l} \ %
    A^{\Lambda *}_{lm}(t) \,%
    X^{\Lambda}_{lm},
\end{equation}
where \(c_{\Lambda}\) 
%is a constant for interaction \(\Lambda\)\CBnote{
is an interaction-dependent constant,
%},  
\(A^{\Lambda}_{lm}(t)\) are time-dependent  components of a spatial spherical tensor, and \(X^{\Lambda}_{lm}\) are  components of a spin or spin-field spherical tensor. 

Spatial spherical tensors may be transformed between arbitrary reference frames $F$ and $G$ by using the Wigner matrices and the Euler angles relating the two frames:
\begin{equation} \label{frame_transformation}
\left[A^{\Lambda}_{lm}\right]^G
    =\sum_{m'=-l}^{+l}
     \left[A^{\Lambda}_{lm'}\right]^F
     D^{(l)}_{m'm}(\Omega_{FG})
\end{equation}
This process may be repeated for a chain of any number of reference frames. Figure~\ref{fig:frame_transformations} depicts the transformations from the principal axis frame of a spin interaction to the laboratory frame. The laboratory-frame spatial components acquire a stochastic time-dependence through the motional modulation of the Euler angles $\WDL(t)$, representing the rotational diffusion of the molecules in solution.

%\begin{enumerate}
%----------
%\item \emph{Direct dipole-dipole coupling.}
\subsubsection{Direct dipole-dipole coupling.}
In the case of the dipole-dipole interaction between spins $I_j$ and $I_k$ ($\Lambda=jk)$, the tensor components $X^{jk}_{2m}$ are equal to the rank-2 spherical tensor spin operators,
\begin{equation}
    X^{jk}_{2m} = T^{jk}_{2m},
\end{equation}
as given in the laboratory frame in table~\ref{tab:spin_tensor_components}. Assuming a rigid molecular geometry, the interaction constant for the dipole-dipole coupling is given by,
\begin{equation}
    c_{jk}=b_{jk}
    =-\left(\frac{\mu_0}{4\pi}\right)\hbar\gamma_j\gamma_k r_{jk}^{-3},
\end{equation}
where $r_{jk}$ is the internuclear distance. In the current case, the \Cth-\Cth internuclear distance of $r_{jk}=122$ pm corresponds to a direct dipole-dipole coupling of $b_{jk}/(2\pi) = -4152.84 \mathrm{~Hz}$.

The principal axis system $P^{jk}$ of the dipole-dipole coupling tensor is aligned such that its z-axis is along the \Cth-\Cth internuclear vector (see figure~\ref{fig:frame_transformations}). In general, the relative orientation of the dipole-dipole principal axis system and the molecular diffusion tensor is defined by an Euler angle triplet $\WPjkD=\{\aPjkD,\bPjkD,\gPjkD\}$, as shown in figure~\ref{fig:frame_transformations}. In the current case, the rod-like geometry of the molecule causes near-coincidence of the principal axis systems of the \Cth-\Cth dipole-dipole coupling and that of the inertia tensor, so that the angle $\bPjkD$ is very small. 

The rank-2 spherical tensor representing the spatial part of the dipole-dipole interaction has the following components in its principal axis frame:
\begin{equation}
   \left[ A^{jk}_{2m} \right]^{P}
    = \sqrt{6}\,\delta_{m0}
\end{equation}
where $\delta_{ab}$ is the Kronecker-delta.
%----------
%\item \emph{Chemical-shift anisotropy}
\subsubsection{Chemical-shift anisotropy.}
%----------
In the case of the chemical shift anisotropy of spin $I_j$ ($\Lambda=j)$, spin-field tensors $X^{j}_{lm}$ of ranks $l=1$ and $l=2$ are formed by coupling the rank-1 spherical tensor spin operators $T^{j}_{1m}$ with the rank-1 spherical components of the external magnetic field~\cite{kowalewski_book_2018}:
\begin{equation}
    X^{j}_{lm} = \sum_{m',m''} C^{l11}_{m m' m''} T^{j}_{1m'} B_{1m''}
\end{equation}
where $C^{l11}_{m m' m''}$ are Clebsch-Gordon coefficients~\cite{VMK_1988}. Explicit expressions for the case $l=2$ are given in the laboratory frame in table ~\ref{tab:spin_tensor_components}.

The magnetic shielding tensors are given in the supplementary material. From these, the \textit{Haeberlen convention}~\cite{haeberlen_high_1976} is used to define the anisotropy and 
%asymmetry 
biaxiality parameters, respectively, as,
\begin{equation}
    \delta^{\mathrm{CSA}} = \delta^{P}_{zz} - \delta^{\mathrm{iso}}
\end{equation}
and,
\begin{equation} \label{biaxiality}
    \eta = \frac{\delta^{P}_{xx}-\delta^{P}_{yy}}{\delta^{\mathrm{CSA}}},
\end{equation}
with tensor components defined by,
\begin{equation}
    |\delta^{P}_{zz} - \delta^{\mathrm{iso}}| \geq |\delta^{P}_{xx} - \delta^{\mathrm{iso}}| \geq |\delta^{P}_{yy} - \delta^{\mathrm{iso}}|.
\end{equation}
Values of these parameters are given in table~\ref{tab:parameters}.

%==============
%\end{enumerate}

\input{tables/tensor_components}
\input{tables/spin_tensor_components}

\subsection{Relaxation Superoperator} \label{relax_superop_section}

The semi-classical relaxation superoperator takes the form,
\begin{equation} \label{relax_superoperator}
    \hat{\Gamma}(t) = 
    - \int_{-\infty}^{0} d\tau {\ }
    \overline{ \hat{\Tilde{H}}_{\mathrm{fluc}}(0) \hat{\Tilde{H}}_{\mathrm{fluc}}(\tau)},
\end{equation}
\noindent where \(\hat{\Tilde{H}}_{\mathrm{fluc}}(t)\) is the fluctuating Hamiltonian commutation superoperator in the interaction representation of the Zeeman Hamiltonian, defined by the transformation,
\begin{equation}
    \hat{\Tilde{H}}_{\mathrm{fluc}}(t) = \mathrm{exp}(i\hat{H}_{\mathrm{Z}}t) \hat{H}_{\mathrm{fluc}}(t) \mathrm{exp}(-i\hat{H}_{\mathrm{Z}}t),
\end{equation}
and the overbar denotes an ensemble average. 

To describe the relaxation effects giving rise to the asymmetric line shapes in figure \ref{fig:full_and_superimposed_spectrum}, the interaction constants and irreducible spherical spin and spatial tensor components in table \ref{tab:tensor_components} are used. By eq.~(\ref{fluctuating_hamiltonian}), the relaxation superoperator may be written as a sum over auto- and cross-correlated mechanisms as,
\begin{equation} \label{linear_relax_superop}
    \begin{split}
        \hat{\Gamma} & = \sum_{\Lambda,\Lambda'} \hat{\Gamma}^{\Lambda\Lambda'} \\
        & = \hat{\Gamma}^{\mathrm{DD}} + \hat{\Gamma}^{\mathrm{CSA}} + \hat{\Gamma}^{\mathrm{DD \times CSA}},
    \end{split}
\end{equation}
%
%where the \((t)\) is dropped for brevity. 
Using eq.~(\ref{H_tensors}), the relaxation superoperator for rank-\(l\) interactions \(\Lambda\) and \(\Lambda'\) becomes,
\begin{equation} \label{relax_superop_tensor_form}
    \hat{\Gamma}^{\Lambda \Lambda'}_{l} = -c^{\Lambda}c^{\Lambda'} \sum_{m} J^{\Lambda\Lambda'}_{lm}(\omega_{0}) \left[\hat{X}^{\Lambda}_{lm}\right]^{L} \left[\hat{X}^{\Lambda'\dagger}_{lm}\right]^{L},
\end{equation}
\noindent with spectral density functions given in our case by,
\begin{equation}
\label{classical_spec_den_func}
\begin{split}
    J^{\Lambda \Lambda'}_{lm}(\omega_{0}) %& = \int^{0}_{-\infty} d\tau \overline{ \Big[A^{\Lambda*}_{lm}(0) \Big]^{L} \Big[A^{\Lambda'}_{l'm'}(\tau) \Big]^{L}} e^{-im'\omega_{0}|\tau|} \\
    & = \left[A^{\Lambda*}_{ln} \right]^{D} \left[A^{\Lambda'}_{ln'} \right]^{D} \int_{-\infty}^{0} d\tau \ G^{ll'}_{mm'nn'} e^{-im'\omega_{0}|\tau|} \\
    & = \sum_{nn'} \frac{(-1)^{n+n'}}{2l+1} \left[A^{\Lambda*}_{ln} \right]^{D} \left[A^{\Lambda'}_{ln'} \right]^{D} \frac{\tau_{\perp}}{1+m^{2}\omega^{2}_{0}\tau^{2}_{\perp}}.
\end{split}
\end{equation}
where \(\left[ A^{\Lambda}_{ln} \right]^{D}\) are \(n^{\mathrm{th}}\)-components of \(l^{\mathrm{th}}\)-rank irreducible spherical tensors in the diffusion frame, and \(\tau_{\perp}\) is the rotation correlation time about an axis perpendicular to the molecular long axis. The components \(\left[ A^{\Lambda}_{ln}\right]^{D}\) are known in the \(P\)-frame and may be expressed in the \(D\)-frame using the transformation in eq.~(\ref{frame_transformation}).

\subsection{Liouvillian} \label{liouvillian_section}

The evolution of the spin ensemble is described by the \textit{Liouville-von Neumann equation}, which may be expressed as,
\begin{equation} \label{LvN_equation}
    \frac{d}{dt} \big|\rho(t)\big) = \hat{\mathcal{L}}(t) \big|\rho(t)\big),
\end{equation}
\noindent where \( \big|\rho(t)\big)\) is the ensemble-averaged density operator, and \(\hat{\mathcal{L}}\) is the Liouvillian, itself given by,
\begin{equation} \label{liouvillian}
    \hat{\mathcal{L}}(t) = -i \hat{H}_{\mathrm{coh}}(t) + \hat{\Gamma}(t),
\end{equation}
\noindent where \(\hat{H}_{\mathrm{coh}}(t)\) is the coherent Hamiltonian commutation superoperator defined by,
\begin{equation}
    \hat{H}_{\mathrm{coh}}(t)\big|Q\big) = \big[ H_{\mathrm{coh}}(t) , Q \big].
\end{equation}
If the Hilbert space of the spin system has dimension \(N_{\mathrm{H}}\), then the corresponding operator (Liouville) space has dimension \(N_{\mathrm{L}} = N^{2}_{\mathrm{H}}\). It follows that the Liouvillian has a set of \(N_{\mathrm{L}}\) eigenvalues and eigenoperators,
\begin{equation} \label{liouvillian_eigenequation}
    \hat{\mathcal{L}} \big|Q_{q}\big) = \Lambda_{q} \big|Q_{q}\big) \qquad q \in \{0,1,\cdots,N_{\mathrm{L}} - 1\},
\end{equation}
\noindent with,
\begin{equation} \label{liouvillian_eigenvalue}
    \Lambda_{q} = -\lambda_{q} + i\omega_{q},
\end{equation}
where \(\lambda_{q}\) and \(\omega_{q}\) are both real. In the case where \(\omega_{q} \neq 0\), the eigenoperators correspond to quantum coherences (QC) which decay with rate constant \(\lambda_{q}\) and oscillate at frequency \(\omega_{q}\). Eigenoperators with real eigenvalues (\(\omega=0\)) represent a particular configuration of spin state populations with decay rate constant \(\lambda_{q}\).

%\MHLnote{I have moved a paragraph here to a later section.}

%In the discussion of linewidths in section \ref{linewidths_section}, we assume that the eigenoperators with complex eigenvalues are close to the corresponding eigenoperators of the coherent Hamiltonian commutation superoperator. This approximation is a good one if the coherence frequencies are well-resolved, i.e. separated by a frequency difference larger than their linewidths. 

\subsection{NMR spectrum} \label{spectrum_section}

\input{tables/eigenoperators}
\subsubsection{Signal} \label{sec:signal}

The signal may be written in terms of the eigenvalues of eq.~(\ref{liouvillian_eigenvalue}) as\cite{ernst_principles_1992},
\begin{equation}\label{eq:s}
    s(t) = \sum_{q} a_{q} e^{\Lambda_{q} t},
\end{equation}
\noindent with \(a_{q}\) the peak amplitude given by,
\begin{equation} \label{peak_amplitude}
    a_{q} = (Q_{\mathrm{obs}}|Q_{q})(Q_{q}|\hat{U}_{\mathrm{exc}}|\rho_{\mathrm{eq}}),
\end{equation}
where \(\hat{U}_{\mathrm{exc}}\) is the total propagator for the excitation sequence and \(\big| \rho_{\mathrm{eq}} \big)\) is the thermal equilibrium density operator. In quadrature detection, \(|Q_{\mathrm{obs}}) \approx -\frac{1}{2}i e^{i\phi_{\mathrm{rec}}} |I^{-})\) with \(\phi_{\mathrm{rec}}\) being the receiver phase. Since the experiment here is a 90\textdegree \ pulse-acquire, we make the approximation,
\begin{equation}
    \hat{U}_{\mathrm{exc}} \big| \rho_{\mathrm{eq}} \big) = \hat{R}_{x}(\pi/2) I_{z} = -I_{y},
\end{equation}
ignoring constant numerical factors.

Non-vanishing peak amplitudes are associated with $(-1)$-quantum eigenoperators $|Q_{q})$, as defined by the eigenequation:
\begin{equation}
    \hat{I}_z |Q_{q}) = -|Q_{q})
\end{equation}
where $\hat{I}_z$ is the commutation superoperator of the angular momentum operator $I_z$. 

In the absence of relaxation, these observable operators are the \((-1)\)-quantum eigenoperators of the commutation superoperator \(\hat{H}_{\mathrm{coh}}\) and are given by elements of the basis,
\begin{equation}\label{eq:eigenops}
\begin{split}
\mathbb{B}_{Q} = \Big\{ & 
\big|| S'_{0} \rangle \langle T'_{+1} |\big),
\big|| T'_{-1} \rangle \langle S'_{0} |\big),
\\ & 
\big|| T'_{0} \rangle \langle T'_{+1} |\big),
\big|| T'_{-1} \rangle \langle T'_{0} |\big) \Big\},.
\end{split}
\end{equation}
which is a subset of the 16-element basis of all outer products of elements in \(\mathbb{B}'_{\mathrm{ST}}\).

In the absence of relaxation, the Liouvillian eigenvalues are purely imaginary, and are given by $\Lambda_{q} = + i\omega_{q}$, where $\omega_q$ are the peak frequencies. These are given in general by 
\begin{equation}
\begin{split}
    \omega_{q} = -(\omega_{r} - \omega_{s}),
\end{split}
\end{equation}
with \( r, s \in \{ S'_{0},T'_{+1},T'_{0},T'_{-1} \} \), as given in table~\ref{tab:coherence_eigen_op_vals}. 

The two eigenoperators corresponding to $(-1)$-quantum coherences between the perturbed triplet states are particularly important in the current context, since these coherences give rise to the two components of the spectral doublet shown in figure~\ref{fig:full_and_superimposed_spectrum}, as can be seen from their amplitudes in table \ref{tab:coherence_eigen_op_vals}. These two eigenoperators are denoted as follows:
\begin{align}
Q_{+} &=
    \big|| T'_{0} \rangle 
    \langle T'_{+1} |\big)
\nonumber\\
Q_{-} &=
    \big|| T'_{-1} \rangle 
    \langle T'_{0} |\big)
\end{align}
The corresponding Liouvillian eigenvalues are as follows:
\begin{align}\label{eq:Lambdapm}
    \Lambda_{\pm} &= -\lambda_{\pm}+ i\omega_{\pm}
\end{align}

In general, the superoperators \(\hat{H}_{\mathrm{coh}}\) and $\hat{\Gamma}$ do not commute. The presence of the relaxation superoperator $\hat{\Gamma}$ may therefore modify both the eigenvalues and the eigenoperators of the Liouvillian $\hat{\mathcal{L}}$. Indeed the modification of the peak frequencies by relaxation effects has been documented in the literature in a different context~\cite{harbison_interference_1993}. In the current case, the eigenvalues of the $(-1)$-quantum eigenoperators are only slightly modified by the relaxation superoperator. This is because the off-diagonal elements of the $(-1)$-quantum Liouvillian block are much smaller than the corresponding eigenvalue differences, as discussed in the Supporting Information. Hence, in the following discussion, we assume that the $(-1)$-quantum eigenoperators of the full Liouvillian, including relaxation, are given to a good approximation by the operators in equation~\ref{eq:eigenops}.

The correspondence between the two triplet-triplet coherences and the NMR spectrum is depicted in figure~\ref{fig:coherences_with_spectrum}.
\input{figures/coherences_with_spectrum}

%\((-1)\)-QC block of the Liouvillian in the \(\mathbb{B}'_{Q}\) basis shows the real part of the eigenvalues to be relatively small perturbations on the coherence frequency, and we assume that the eigenoperators with complex eigenvalues are close to the corresponding eigenoperators of the coherent Hamiltonian commutation superoperator. \JWnote{full liouvillian matrix now in the SM. The full matrix has been diagnalised in mathematica and I am working on presenting it. The basis operators are a large combination of ket-bra products.}
%This matrix is given in the supplementary information. As such, by ignoring relaxation here, we may focus on the positions and amplitudes of the peak and assign them to coherences.
%\MHLnote{I have replaced the paragraph by a briefer reference to the SI. }
%
%\MHLnote{The paragraph below is best placed in the ``results" section. It's better to keep the theory general, and only substitute in specific numbers in ``results"} \JWnote{amplitudes moved to Results.}

%------------------------
\subsubsection{Frequencies} \label{sec:frequencies}
The coherence frequencies are given by the imaginary parts of the Liouvillian eigenvalues. As shown in the Supporting Information, the off-diagonal parts of the $(-1)$-quantum Liouvillian block may be ignored. With this approximation, the coherence frequencies are as specified in Table~\ref{tab:coherence_eigen_op_vals}. The frequencies of the two triplet-triplet coherences are given by
\begin{equation} \label{eq:frequencies}
\omega_{\pm} =
\frac{1}{2} \big(
\omega_{\Sigma} \pm \omega_{J} \mp \omega_{e}
\big)
\end{equation}
%
%\MHLnote{Give an approximate formula for the frequency splitting between the two peaks, in the near-equivalence limit.}
%
The splitting between the two inner peaks is given by,
\begin{align}
    \Delta\omega &= 
    \omega_{-} - \omega_{+} 
= \omega_e - \omega_J 
\simeq 
 \frac{\omega_\Delta^2}{2\omega_J}.
\end{align}
where the approximation applies to the near-equivalence regime.
%------------------------
\subsubsection{Linewidths} \label{sec:linewidths}
Since the off-diagonal elements of \(\hat\Gamma\) are small in the basis \(\mathbb{B}_{Q}\), relative to the corresponding differences in the diagonal elements, the real parts of the Liouvillian eigenvalues are given by
\begin{equation} \label{decoherence_rate}
%- \lambda_q =
\mathrm{Re}(\Lambda_q) \simeq 
\frac{\big( Q_{q} \big| \hat{\Gamma} \big| Q_{q} \big)}{\big( Q_{q} \big| Q_{q} \big)}
\end{equation}
where the Liouville bracket is defined by~\cite{jeener_superoperators_1982},
\begin{equation}
    \big( Q_{q} \big| Q_{q'} \big) = \mathrm{Tr}\big\{ Q^{\dagger}_{q} Q_{q'} \big\}.
\end{equation}
The real positive quantities $\lambda_q = -\mathrm{Re}(\Lambda_q)$ may be interpreted as the coherence decay rate constants for the eigenoperators $\big| Q_{q} \big)$. After Fourier transformation of the NMR spectrum, the peak associated with the eigenoperator $\big| Q_{q} \big)$ has amplitude $a_q$, centre frequency $\omega_q$, and has a Lorentzian shape with half-width-at-half-height equal to $\lambda_q$, in units of $\mathrm{rad \ s^{-1}}$. Its full-width-at-half-height is given by $\lambda_q/\pi$ in units of Hz.

The relaxation superoperator $\hat\Gamma$ may be written as a sum of auto-correlation terms for the DD and CSA interactions, and a DD$\times$CSA cross-correlation term (eq. (\ref{linear_relax_superop})). The coherence decay rate constants $\lambda_q$ may therefore be written as a superposition of terms:
\begin{equation}
\label{eq:lamqsuperpos}
\lambda_q =
    \lambda_q^\mathrm{DD}+
    \lambda_q^\mathrm{CSA}+
    \lambda_q^\mathrm{DD\times CSA}.
\end{equation}
%
%\green{To simplify relaxation rate expressions, we note that if \One had perfect cylindrical symmetry about the labelled spins, then by eq. (\ref{biaxiality}) \(\eta_{j(k)}\) would vanish. Indeed, the computed \(\eta_{j(k)}\) are very small for our system (see table~\ref{tab:parameters}). Making the approximation that \(\eta_j \simeq \eta_k \simeq 0\), all components of the spatial tensor associated with the CSA interaction vanish except for \(\left[ A^{\mathrm{CSA}}_{20} \right]^{P} = \sqrt{3/2}\,\delta^{\mathrm{CSA}}\), and the transformation in eq.~(\ref{frame_transformation}) reduces to,}
%
The computed CSA biaxiality parameters $\eta$ are very small for both \Cth sites of the system \One (see table~\ref{tab:parameters}). Making the approximation that \(\eta_j \simeq \eta_k \simeq 0\), all components of the spatial tensor associated with the CSA interaction vanish except for \(\left[ A^{\mathrm{CSA}}_{20} \right]^{P} = \sqrt{3/2}\,\delta^{\mathrm{CSA}}\), and the transformation in eq.~(\ref{frame_transformation}) reduces to,
\begin{equation}
\begin{split}
    \left[ A^{\mathrm{CSA}}_{2n} \right]^{D}  & = \left[ A^{\mathrm{CSA}}_{20} \right]^{P}D^{(2)}_{0n}(\WPjD) \\
    & = \left[ A^{CSA}_{20} \right]^{P},
\end{split}
\end{equation}
where the last line is obtained by noting that the $P$- and $D$-frames are coincident, and all Euler angles may be set to zero.

For the two triplet-triplet coherences, each
term in eq. (\ref{eq:lamqsuperpos}) is given by,
\begin{equation} \label{r2_DD}
    \lambda^{\mathrm{DD}}_{\pm} = \frac{3}{20} b^{2}_{jk} \tau_{\perp} \bigg( 3 +  \frac{3}{1 +  \omega^{2}_{0}\tau^{2}_{\perp}} + \frac{2}{1 + 4\omega^{2}_{0}\tau^{2}_{\perp}} \bigg),
\end{equation}
\begin{equation} \label{r2_CSA}
\begin{split}
%    R^{\mathrm{CSA}(\pm)}_{2} =
    \lambda^{\mathrm{CSA}}_{\pm} =\frac{1}{20} & \omega^{2}_{0} \tau_{\perp} \bigg\{ \left( \big[\delta^{\mathrm{CSA}}_{j}\big]^{2} + \big[\delta^{\mathrm{CSA}}_{k}\big]^{2} \right) \frac{5 + 2\omega^{2}_{0}\tau^{2}_{\perp}}{1 + \omega^{2}_{0}\tau^{2}_{\perp}} \\ & + \delta^{\mathrm{CSA}}_{j} \delta^{\mathrm{CSA}}_{k}  \frac{3}{1 + \omega^{2}_{0}\tau^{2}_{\perp}} \bigg\},
\end{split}
\end{equation}
and,
\begin{equation} \label{r2_DDxCSA}
    \lambda^{\mathrm{DD\times CSA}}_{\pm}
%    R^{\mathrm{DD \times CSA}(\pm)}_{2} 
    = \pm \frac{3}{20} \omega_{0} b_{jk} \tau_{\perp} \Big(\delta^{\mathrm{CSA}}_{j} + \delta^{\mathrm{CSA}}_{k}\Big) \frac{3 + 2 \omega^{2}_{0}\tau^{2}_{\perp}}{1 +  \omega^{2}_{0}\tau^{2}_{\perp}},
\end{equation}
%
%where the 
%superscript 
%subscript
%\((\pm)\) denotes \(\big| Q_{q} \big) = \big| | T_{0} \rangle \langle T_{+1} |\big)\) \textit{or} \(\big| | T_{-1} \rangle \langle T_{0} |\big)\), respectively, and the approximation \(\mathbb{B}'_{\mathrm{ST}} \approx \mathbb{B}_{\mathrm{ST}}\) is also used since \(\theta \approx 0\).

Equations~(\ref{r2_DD}) - (\ref{r2_DDxCSA}) depend on the correlation time \(\tau_{\perp}\) for molecular rotation around an axis perpendicular to the long axis of the molecule. Rotational diffusion \emph{around} the molecular long axis does not modulate the spin interactions,
%relaxation interactions in our system, 
under the approximation of a rigid symmetric top undergoing rotational diffusion, and does not lead to spin relaxation.

%\MHLnote{It would be good to give equations here for $\lambda_\pm$ in the two limiting cases: extreme narrowing, and the long-correlation-time limit. The two limits may be stated without referring to our specific experimental case. Discussion of our specific case is best placed in ``results", where we compare the analysis with the experimental data.} \JWnote{$\lambda_{\pm}$ given below in limiting cases, with fig. \ref{fig:lambda_vs_B0_plots} plotting these against B0. Amplitudes, $\lambda_{\pm}$ and $\omega_{\pm}$ used for Lorentzian plot moved to results.}

In the current case, the chemical shift anisotropies of the two spins are very similar, allowing the simplification $\delta^{\mathrm{CSA}}\simeq \delta^{\mathrm{CSA}}_{j}\simeq \delta^{\mathrm{CSA}}_{k}$. 

The limiting regimes of the correlation time $\tau_\perp$ are as follows:
\begin{enumerate}
    \item 
In the \emph{extreme narrowing limit}, $|\omega_{0}\tau_{\perp} |\ll 1$,
%within 0.1 ppm of each other, and \(|\tau_{\perp} \omega_{0}|^{2} = 7.84 \times 10^{-3}\). Making the approximation $\delta^{\mathrm{CSA}}\simeq \delta^{\mathrm{CSA}}_{j}\simeq \delta^{\mathrm{CSA}}_{k}$ and invoking the extreme narrowing limit, 
eq. (\ref{eq:lamqsuperpos}) may be written,
\begin{equation} \label{approximate_rates}
    \lambda_{\pm} \simeq
    \frac{3}{10} \left( 4 b_{jk} \pm 3 \omega_{0} \delta^{\mathrm{CSA}} \right) b_{jk} \tau_{\perp} + \lambda^{\mathrm{CSA}},
%    \lambda^{\mathrm{CSA}}_{\pm},
\end{equation}
where the CSA-induced decay rate constant \(\lambda^{\mathrm{CSA}}\) is given by,
\begin{equation}
\label{eq:lambdaCSA}
    \lambda^{\mathrm{CSA}} \simeq \frac{13}{20} \omega^{2}_{0} \left[ \delta^{\mathrm{CSA}} \right]^{2} \tau_{\perp}.
\end{equation}
The field-dependence of the two rate constants $\lambda_\pm$ is illustrated in figure~\ref{fig:lambda_vs_B0_plots}a. The decay rate constant $\lambda_{+}$ is minimized at a  magnetic field such that $\left| 4 b_{jk} \right| = \left| 3 \omega_{0}\delta^{\mathrm{CSA}}\right|$, in which case the first term in eq. (\ref{approximate_rates}) cancels out. At this field, the dipole-dipole contribution to the decay rate constant vanishes, and $\lambda_{+}$ becomes equal to the limiting CSA relaxation rate constant $\lambda^{\mathrm{CSA}}$ (eq. (\ref{eq:lambdaCSA})). The decay rate constant $\lambda_{+}$, on the other hand, increases monotonically with increasing magnetic field. 
%
%with the \(j\) and \(k\) subscripts dropped since the anisotopy is equivalent for both spins.
%
%\MHLnote{ok, it turns out that in the long $\tau_\perp$ limit, in the appropriate field, the CSA and the DD do cancel out for one of the doublet components, as in TROSY. But this does not happen in the extreme narrowing limit. This is a good topic for the discussion.}

\item %
In the \emph{long correlation time limit}, \(|\omega_{0} \tau_{\perp}| \gg 1\), eq.~(\ref{eq:lamqsuperpos}) may be written as,
%
%This is utilised in biomolecular TROSY NMR and used to explain the cancellation of DD and CSA mechanisms. In this limit, we may write eq.~(\ref{eq:lamqsuperpos}) for the triplet-triplet coherences as,
%
\begin{equation} \label{long_tau_limit}
    \lambda_{\pm} \simeq \frac{1}{20} \left( 3 b_{jk} \pm 2 \omega_{0} \delta^{\mathrm{CSA}} \right)^{2} \tau_{\perp}.
\end{equation}
The field-dependence of the two rate constants $\lambda_\pm$ is illustrated in figure~\ref{fig:lambda_vs_B0_plots}b. In this regime, the linewidth parameter $\lambda_{+}$ goes to zero at a magnetic field such that $ \left| 3 b_{jk} \right| = \left| 2 \omega_{0} \delta^{\mathrm{CSA}} \right| $. The strong narrowing of one of the two doublet components resembles the TROSY effects exploited in biomolecular NMR~\cite{pervushin_TROSY_1997,tugarinov_MeTROSY_2003}.

\end{enumerate}

%
%For eq. (\ref{long_tau_limit}), a 4 T field completely cancels the DD and CSA interactions for the \(Q_{+}\) coherence, and the peak-width vanishes. This is illustrated in fig. \ref{fig:lambda_vs_B0_plots}.

\input{figures/lambda_vs_B0_plots}
\section{Results}

%Peak assignment is determined by the theory in sections \ref{sec:signal}, \ref{sec:frequencies}, and \ref{sec:linewidths}, and analytical parameters obtained are given in table \ref{tab:analytical_results}.

Using eq. (\ref{peak_amplitude}), the peak amplitudes associated with the \((-1)\)-quantum singlet-triplet coherences are given by \( \propto \mathrm{sin}^{2}(\theta/2) \), while those associated with the \((-1)\)-quantum triplet-triplet coherences are given by \( \propto \mathrm{cos}^{2}(\theta/2) \). In the current case, the singlet-triplet mixing angle is small (\(\theta = -0.0750=-4.30^\circ\)), and the amplitudes are,
\begin{equation} \label{numerical_amplitudes}
    \begin{split}
        a_{S_{0}'\xrightarrow{}T_{+1}'} =  a_{T_{-1}'\xrightarrow{}S_{0}'} & \simeq 0.686 \times 10^{-3} \\
%\blue{a_{\pm} = {\ }{\ } }
        a_{T_{0}'\xrightarrow{}T_{+1}'} =  a_{T_{-1}'\xrightarrow{}T_{0}'} & \simeq 0.499,
    \end{split}
\end{equation}
\noindent with the sum over all amplitudes equal to 1. The spectrum is therefore dominated by the strong peaks from the two triplet-triplet coherences. 
%This indicates that the peaks in fig. \ref{fig:full_and_superimposed_spectrum} correspond to the \((-1)\)-QC in the triplet manifold, described by the operators, \(Q_{+} = \big|| T'_{0} \rangle \langle T'_{+1} |\big),\) and \(Q_{-} = \big|| T'_{-1} \rangle \langle T'_{0} |\big)\), while those associated with \(\big|| S'_{0} \rangle \langle T'_{+1} |\big),\) and \(\big|| T'_{-1} \rangle \langle S'_{0} |\big)\) are vanishingly small in the current case.

From eqs.~(\ref{eq:we}) and (\ref{eq:frequencies}), the frequency \(\omega_{+}\) is less than \(\omega_{-}\). This indicates that the left peak of the doublet is associated with the \(Q_{+}\) coherence, while the right-hand peak is associated with the \(Q_{-}\) coherence, after taking into account the sign of the Larmor frequency~\cite{levitt_signs_1997}. This assignment is shown in figure~\ref{fig:coherences_with_spectrum}. The splitting between the peaks is given by $\Delta\omega/(2\pi)=0.60\mathrm{\,Hz}$.

%In the rotating frame, we may Set \(\omega_{\Sigma} = 0\). We then have,
%
%\begin{equation}
%   \omega_{\pm} = \mp 1.90 \ \mathrm{rad \ s^{-1}}.
%\end{equation}
%
From eqs.~(\ref{r2_DD})-(\ref{r2_DDxCSA}), since \(b_{jk}\), \(\omega_{0}\), \(\delta^{\mathrm{CSA}}_{j}\) and \(\delta^{\mathrm{CSA}}_{k}\) are all negative, we see that the cross-correlation contributions reduce the value of \(\lambda_{+}\) while increasing the value of \(\lambda_{-}\). 
%This is consistent with table~\ref{tab:coherence_eigen_op_vals}. 
For the experimental parameters, the coherence decay rate constants are given by $\lambda_{+} \simeq 0.583 \ \mathrm{\ s^{-1}}$ and $\lambda_{-} \simeq 1.190 \ \mathrm{\ s^{-1}}$. These correspond to full peakwidths at half-height of 0.186~Hz and 0.379~Hz, for the left-hand and right-hand doublet components, respectively.

%
%\begin{align}
%    \lambda_{+} &\simeq 0.583 \ \mathrm{\ s^{-1}},
%\nonumber\\
 %   \lambda_{-} &\simeq 1.19 \ \mathrm{\ s^{-1}}.
%\end{align}
%
%
%Taken together, the spectral peaks may be assigned with confidence; the narrowed left peak and the broadened right peak are derived from the \(Q_{+}\) and \(Q_{-}\) coherences, respectively. This is illustrated in fig.~\ref{fig:coherences_with_spectrum}.

%These analytical results based on the theory we present are gathered in table \ref{tab:analytical_results} for the inner doublet. The superposition of absorption Lorentzians,
The green curve in figure~\ref{fig:full_and_superimposed_spectrum} is a plot of the analytical spectral function
\begin{equation} \label{lorentzian}
\begin{split}
    S(\omega)
%    \mathcal{A}\left( \omega; a_{\pm}, \lambda_{\pm}, \omega_{\pm} \right) 
    = \ & a_{+} \frac{\lambda_{+}}{\lambda^{2}_{+} + \left( \omega - \omega_{+} \right)^{2}} \\ & + a_{-} \frac{\lambda_{-}}{\lambda^{2}_{-} + \left( \omega - \omega_{-} \right)^{2}},
\end{split}
\end{equation}
using the parameters in table~\ref{tab:analytical_results}. There is good agreement with the experimental \Cth NMR spectrum (black).

The blue curve in figure~\ref{fig:full_and_superimposed_spectrum} shows the result of a numerical calculation using \textit{SpinDynamica}~\cite{bengs_spindynamica_2018} in which the full Liouvillian is diagonalized. There is good qualitative agreement between the numerical simulations, the analytical theory and the experimental result. The minor differences between the \emph{SpinDynamica} simulation and the analytical theory may be attributed to the neglect of the off-diagonal Liouvillian elements in the analytical theory (see discussion after eq. (\ref{eq:Lambdapm})). 

%\ToHere
%
%as a function of these parameters is plotted alongside the experimental spectrum in fig. \ref{fig:full_and_superimposed_spectrum}. We see there is also excellent qualitative agreement between theory and experiment in the analytical case.

%The 90\textdegree \ pulse-acquire spectrum was simulated in \textit{SpinDynamica}~\cite{bengs_spindynamica_2018} using the parameters in table \ref{tab:parameters}, in which the full Liouvillian is diagonalised, and is plotted in fig. \ref{fig:full_and_superimposed_spectrum}b alongside the experimental spectrum. There is good qualitative agreement between theory and experiment. Since the relaxation mechanisms considered in the simulation were restricted to DD and CSA interactions, as well as cross-correlation between the two, this confirms that the asymmetric lineshape in fig. \ref{fig:full_and_superimposed_spectrum}b is a consequence of strong correlation between these mechanisms.

\input{tables/analytical_results}

\section{Conclusions}
The results and theory reported here show that cross-correlated relaxation can have a strong effect on the NMR spectra of homonuclear spin-1/2 pairs in the near-equivalence regime. This has strong relevance to NMR experiments on long-lived states, which are often performed on spin systems of this kind~\cite{pileio_long-lived_2020,MHL-LLS-ARPC_2012,stevanato_nuclear_2015,Levitt-LLSS_2019}. 

In a following paper, we explore the influence of cross-correlated relaxation on the \emph{longitudinal} relaxation of spin systems of this kind, including the relaxation of long-lived states.

\section*{Acknowledgements}

We thank Laurynas Dagys for insightful discussions on both experiment and theory. This research was supported by the European Research Council (grant 786707-FunMagResBeacons) and EPSRC-UK (grants EP/P030491/1, EP/P009980/1). We also acknowledge the IRIDIS High Performance Computing Facility, and associated support services at the University of Southampton.

\section*{Author Declarations}

\subsection*{Conflict of interest}

The authors have no conflicts to disclose.

\section*{Data Availability Statement}

The data that support the findings of this study are available from the authors at reasonable request.

\section*{Supplementary Information}

The derivation of the relaxation superoperator in more detail, and the synthesis details of \One are in a document offered alongside the paper.

\bibliography{main.bbl}

%==========================

\end{document}

\title{Cross-correlation effects in the solution NMR %spectra 
of near-equivalent
%$\mathrm{{}^{13}C}$ 
spin-1/2 pairs: Supplementary Material}

\date{\today}%

\author[1]{James W. Whipham}
\author[1]{Gamal A. I. Moustafa}
\author[1]{Mohamed Sabba}
\author[1]{Weidong Gong}
\author[1]{Christian Bengs}
\author[1*]{Malcolm H. Levitt}

\affil[*]{mhl@soton.ac.uk}

\affil[1]{Department of Chemistry, University of Southampton, SO17 1BJ, UK}%

\maketitle

\newpage

\section{700 MHz Spectrum}

\begin{figure}[tb] \label{relax_superop_plot}
\centering
\includegraphics[width=0.9\linewidth]{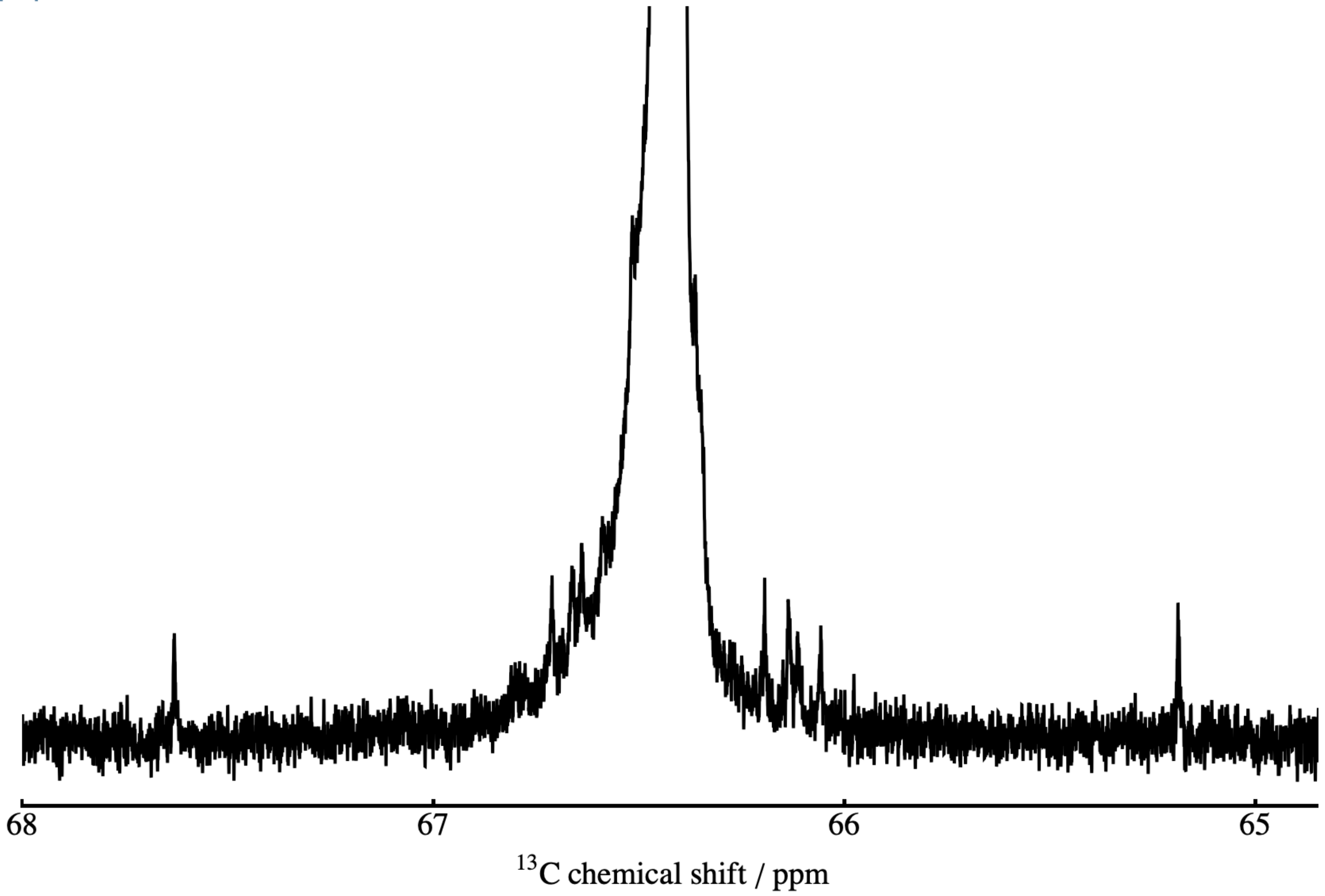}
\caption{
A portion surrounding the main doublet of the 700 MHz \(90^{\circ}\) pulse-acquire spectrum, showing the weak transitions at 65.19 and 67.63 ppm.
}
\label{fig:700_spectrum} 
\end{figure}

A portion of the \(90^\circ\) pulse-acquire spectrum taken on a 700 MHz (16.4 T) spectrometer is shown in fig. \ref{fig:700_spectrum}, showing the weak outer-transitions at 65.19 and 67.63 ppm. Using table IV in the main text, we see that these peaks are associated with the \(\big| |S'_{0} \rangle \langle T'_{+1} | \big)\) and \(\big| |T'_{-1} \rangle \langle S'_{0} | \big)\) \((-1)\)-quantum coherences, respectively.

The 700 MHz spectrum was also used to determine the isotropic \(J\)-coupling between the labelled nuclei as 214.15 Hz, by measuring the splitting between an outer peak and the associated inner-peak.

\section{Relaxation Superoperator}

\subsection{Derivation}

A spin Hamiltonian for interaction \(\Lambda\) may be written as a tensor product between a time-dependent spatial tensor and time-independent spin tensor. A convenient set of tensors to use are the \textit{irreducible spherical tensor operators}, since these are eigenoperators of the Zeeman Hamiltonian commutation superoperator, leading to a simple expression in the corresponding interaction representation. For a rank-\textit{l} interaction, we have,~\cite{smith_hamiltonians_1992}
\begin{equation} \label{H_tensor_form}
    H^{\Lambda}_{l}(t) = c^{\Lambda} \sum_{m=-l}^{+l} A^{\Lambda*}_{lm}(t) X^{\Lambda}_{lm},
\end{equation}
\noindent where \(c^{\Lambda}\) is a real constant specific to interaction \(\Lambda\). Being Hermitian, we may also write,
\begin{equation} \label{H_hermitian}
\begin{split}
    H^{\Lambda}_{l}(t) & = H^{\Lambda \dagger}_{l} \\ & = c^{\Lambda} \sum_{m=-l}^{+l} A^{\Lambda}_{lm}(t) X^{\Lambda \dagger}_{lm}.
\end{split}
\end{equation}
Useful relations for components \(A_{lm}\) and \(X_{lm}\) are,
\begin{equation}
    A^{*}_{lm} = (-1)^{m} A_{l-m},
\end{equation}
\noindent and,
\begin{equation}
    X^{\dagger}_{lm} = (-1)^{m} X_{l-m}.
\end{equation}
\noindent Then, using the eigenoperator relation,
\begin{equation}
    \big[I_{z},X_{lm}\big] = mX_{lm},
\end{equation}
\noindent the tensor components may be written in the interaction representation of the Zeeman interaction as,
\begin{equation}
\begin{split}
    \Tilde{X}_{lm}(t) & = e^{i\hat{H}_{\mathrm{Z}}t} X_{lm}\\
    & = X_{lm}e^{im\omega_{0}t},
\end{split}
\end{equation}
\noindent where \(\hat{H}_{\mathrm{Z}}\) is the Zeeman Hamiltonian commutation superoperator, \(\hat{H}_{\mathrm{Z}} = \big[H_{\mathrm{Z}}, \cdots \big] = \omega_{0}\big[I_{\mathrm{z}}, \cdots \big]\).
In a \textit{Wangsness-Bloch-Redfield} (WBR) formalism, the relaxation superoperator takes the form,
\begin{equation}
    \hat{\Gamma} = - \frac{1}{2} \int^{+\infty}_{-\infty} d\tau \overline{ \hat{\Tilde{H}}_{\mathrm{fluc}}(t) , \hat{\Tilde{H}}_{\mathrm{fluc}}(t+\tau)}.
\end{equation}
\noindent From here, we simplify the expression. First, we have the liberty of setting \(t=0\) by assuming the noise in the system is stationary. This removes an exponential term in the interaction representation. Also, by neglecting the small dynamic frequency shifts, the correlation function has time-reversal symmetry and the integral may be taken from \(-\infty\) to \(0\) while introducing a factor of two. We now write,  

\begin{equation}
    \hat{\Gamma} = -  \int^{0}_{-\infty} d\tau \overline{ \hat{\Tilde{H}}_{\mathrm{fluc}}(0) \hat{\Tilde{H}}_{\mathrm{fluc}}(\tau)}.
\end{equation}
\noindent \(\hat{\Gamma}\) may be decomposed as a sum of auto- and cross-correlated relaxation superoperators. Simply,
\begin{equation}
    \hat{\Gamma} = \sum_{\Lambda,\Lambda'} \sum_{l,l'} \hat{\Gamma}^{\Lambda \Lambda'}_{ll'},
\end{equation}
\noindent and using eq. (\ref{H_tensor_form}) while utilising (\ref{H_hermitian}), \(\hat{\Gamma}^{\Lambda \Lambda'}_{ll'}\) takes the form,
\begin{equation}
    \hat{\Gamma}^{\Lambda\Lambda'}_{ll'} = -c^{\Lambda}c^{\Lambda'} \sum_{mm'} J^{\Lambda\Lambda'}_{ll'mm'}(\omega_{0}) \Big[\hat{X}^{\Lambda}_{lm}\Big]^{L} \Big[\hat{X}^{\Lambda'\dagger}_{l'm'}\Big]^{L},
\end{equation}
\noindent where the spectral density function is given by,
\begin{equation} \label{spectral_density_function}
 J^{\Lambda\Lambda'}_{ll'mm'}(\omega_{0}) = \mathrm{Re} \int^{0}_{-\infty} d\tau \overline{ \Big[A^{\Lambda*}_{lm}(0) \Big]^{L} \Big[A^{\Lambda'}_{l'm'}(\tau) \Big]^{L}} e^{-im'\omega_{0}|\tau|}.
\end{equation}
\noindent Here, the square brackets with the superscript \(L\) denote the laboratory frame. This is important since the spatial functions and spin tensors must be expressed in the same frame to legitimise the Hamiltonian they are associated with. However, the spatial functions are known in the molecule-fixed principal axis (\(P\)-) frame of the interaction in question, whereas the spin tensors are known in the space-fixed \(L\)-frame. Thus, we transform the spatial functions in the \(L\)-frame as a linear combination of functions in the diffusion (\(D\)-) frame using the properties of Wigner functions, before themselves being transformed as a linear combination of functions in the \(P\)-frame.

The relation to use is,
\begin{equation} \label{wigner_rotation_relation}
    \Big[A_{lm}(t)\Big]^{L} = \sum_{n} \Big[A_{ln}\Big]^{D} D^{(l)}_{nm}(\Omega_{DL}(t)),
\end{equation}
\noindent and the ensemble-averaged product in eq. (\ref{spectral_density_function}) for the system becomes,
\begin{equation}
\begin{split}
    \overline{ \Big[A^{\Lambda*}_{lm}(0) \Big]^{L} \Big[A^{\Lambda'}_{l'm'}(\tau) \Big]^{L}} = \sum_{nn'} & \overline{D^{(l)}_{mn}(\Omega_{LD}(0)) D^{(l')*}_{m'n'}(\Omega_{LD}(\tau))} \\
    & \times \Big[A^{\Lambda*}_{ln}\Big]^{D} \Big[A^{\Lambda'}_{l'n'}\Big]^{D},
\end{split}
\end{equation}
\noindent where the relation \(D^{(l)}_{nm}(\Omega)=D^{(l)*}_{mn}(\Omega^{-1})\) is used, and \(\Omega^{-1}_{DL} \equiv \Omega_{LD}\). \noindent The time-dependence of the spatial functions in the \(L\)-frame has been absorbed into the Wigner functions, since the \(D\)-frame is molecule-fixed. We then define the time-correlation function as that between the Wigner functions only, as,
\begin{equation} \label{time_corr_function}
\begin{split}
    G^{ll'}_{mm'nn'}(\tau) & = \overline{D^{(l)}_{mn}(\Omega_{LD}(0)) D^{(l')*}_{m'n'}( \Omega_{LD}(\tau))} \\
    & = \int \int d\Omega(0) d\Omega(\tau) D^{(l)}_{mn}(\Omega(0)) D^{(l')*}_{m'n'}(\Omega(\tau)) \times P(\Omega(0)) P(\Omega(\tau)|\Omega(0)),
\end{split}
\end{equation}
\noindent where \(\Omega^{-1}_{DL}\) have been denoted simply by \(\Omega\) in the second line for brevity, \(P(\Omega(0)) = 1/(8\pi^{2})\) and is the probability that the molecule hosting the spin system will be at orientation \(\Omega\) at time \(t=0\), and \(P(\Omega(\tau)|\Omega(0))\) is the \textit{conditional probability} that the molecule will be at orientation \(\Omega(\tau)\) at time \(t' = t+\tau = \tau\) given that it was at orientation \(\Omega(0)\) at time \(t=0\). From here, the notation and derivation is in similar vein of Huntress~\cite{huntress_effects_1968,huntress_study_1970}.

%Huntress showed that in the frame which diagonalises the diffusion tensor (i.e. the \(D\)-frame), the conditional probability may be written as a series of inner products between asymmetric rigid-rotor eigenfunctions at times \(t=0\) and \(\tau\). That is,

The time-derivative of the probability that the molecule will be at orientation \( \Omega(\tau) \) at time \(\tau\), in the limit of a rigid molecule reorienting in random steps of small angular displacement, is given by the \textit{Favro equation}~\cite{favro_theory_1960},
\begin{equation} \label{favro}
    \frac{\partial}{\partial \tau} P(\Omega(\tau)) = - H_{\mathrm{rot-diff}} P(\Omega(\tau)),
\end{equation}
\noindent where \( H_{\mathrm{rot-diff}} \) is the \textit{rotational-diffusion Hamiltonian}, which may be written in the form,
\begin{equation} \label{diffusion_operator}
    H_{\mathrm{rot-diff}} = \mathbf{L} \bullet \mathbf{D} \bullet \mathbf{L}
\end{equation}
\noindent where \( \mathbf{L} \) and \( \mathbf{D} \) are the quantum mechanical angular momentum operator and diffusion tensor, respectively. That is, the diffusion tensor describes the spatial aspect of the Hamiltonian in this case.

Favro shows that the solution to equation (\ref{favro}) is~\cite{favro_theory_1960},
\begin{equation} \label{probability_at_tau}
    P(\Omega(\tau)) = \int d\Omega(0) P(\Omega(0)) P(\Omega(0)|\Omega(\tau)),
\end{equation}
\noindent for which the conditional probability is,
\begin{equation} \label{conditional_probability}
    P(\Omega(0)|\Omega(\tau)) 
    = \sum_{\nu} \psi^{*}_{\nu}(\Omega(0)) \psi_{\nu}(\Omega(\tau)) e^{-E_{\nu} \tau},
\end{equation}
noindent where \( \psi(\Omega) \) are eigenfunctions of \( H_{\mathrm{rot-diff}} \) with the associated eigenvalue \( E_{\nu} \). This solution is reliant on the boundary condition \( P(\Omega(0)|\Omega(\tau\rightarrow0)) = \delta(\Omega(0),\Omega(\tau)) \), where \( \delta(\Omega(0),\Omega(\tau)) \) is the Dirac delta function.

If eq. (\ref{diffusion_operator}) is written in the \(D\)-frame, it takes the form \( H_{\mathrm{rot-diff}} = \sum_{i} D_{i} L^{2}_{i} \) in which \( i \in \{ x, y, z \} \) and \(L_{i}\) have become the Cartesian angular momentum operators. This takes the form of the rigid-rotor Hamiltonian when considering the substitution \( D_{i} \rightarrow \hbar^{2} / 2I_{i} \), where \( I_{i} \) is the moment of inertia about principal axis \( i \). Thus, (\ref{conditional_probability}) may be expanded in asymmetric-rotor eigenfunctions,
\begin{equation}
    P(\Omega(\tau)|\Omega(0)) = \sum_{\nu,J,M} \psi^{J*}_{\nu,M}(0) \psi^{J}_{\nu,M}(\tau) e^{-E^{J}_{\nu}\tau},
\end{equation}
\noindent where,
\begin{equation} \label{asymmetric_rotor_eigenfunctions}
    \psi^{J}_{\nu,M}(t) = \sum_{K} a^{J}_{\nu,K} \phi^{J}_{K,M}
\end{equation}
\noindent and \(\phi^{J}_{K,M}\) are symmetric-rotor eigenfunctions which may be expressed in terms of Wigner functions and take the form,
\begin{equation} \label{symmetric_rotor_eigenfunctions}
    \phi^{J}_{K,M} = (-1)^{M-K}\sqrt{\frac{2J+1}{8\pi^{2}}} D^{(J)}_{-M-K}(\Omega).
\end{equation}
We then have all we need to obtain the time-correlation function and subsequently the spectral density function and relaxation superoperator. Inserting these eigenfunctions and probabilities into (\ref{time_corr_function}) and setting \(\Omega \equiv \Omega_{LD}\) in (\ref{symmetric_rotor_eigenfunctions}),
\begin{equation}
\begin{split}
    G^{ll'}_{mm'nn'}(\tau) = &  \frac{1}{8\pi^{2}} \sum_{\nu,J,M} (2J+1) e^{-E^{J}_{\nu}\tau} \\ & \times \int \int d\Omega_{LD}(0) d\Omega_{LD}(\tau) D^{(l)}_{mn}(\Omega_{LD}(0)) D^{(l')*}_{m'n'}(\Omega_{LD}(\tau)) \\
    & \times \bigg\{ \sum_{K} a^{J*}_{\nu,K} (-1)^{M-K} D^{(J)*}_{-M-K}(\Omega_{LD}(0)) \bigg\} \\
    & \times \bigg\{ \sum_{K'} a^{J}_{\nu,K'} (-1)^{M-K'} D^{(J)}_{-M-K'}(\Omega_{LD}(\tau)) \bigg\},
\end{split}
\end{equation}
\noindent and rearranging the expression, while using the orthogonality relation,
\begin{equation}
    \int d\Omega D^{j}_{pq}(\Omega) D^{j'}_{p'q'}(\Omega) = \frac{8\pi^{2}}{2j+1} \delta_{jj'} \delta_{pp'} \delta_{qq'},
\end{equation}
\noindent the time correlation function simplifies to,
\begin{equation}
    \begin{split}
        G^{l}_{mnn'}(\tau) & = \delta_{ll'}\delta_{mm'} G^{ll'}_{mm'nn'} (\tau) \\
        & = \frac{(-1)^{n+n'}}{2l+1} \sum_{\nu} a^{l*}_{\nu,n} a^{l}_{\nu,-n'}e^{-E^{l}_{\nu}\tau},
    \end{split}
\end{equation}
\noindent where \(l=l'=J\), \(m=m'=-M\), \(n=-K\) and \(n'=-K'\).

From this, the spectral density becomes,
\begin{equation}
    \begin{split}
        J^{\Lambda \Lambda'}_{lm}(\omega_{0}) & = \delta_{ll'}\delta_{mm'} J^{\Lambda \Lambda'}_{ll'mm'}(\omega_{0}) \\
        & = \mathrm{Re} \int_{-\infty}^{0} d\tau \overline{ \Big[A^{\Lambda*}_{lm}(0) \Big]^{L} \Big[A^{\Lambda'}_{lm}(\tau) \Big]^{L} } e^{-im\omega_{0}|\tau|} \\
        & 
        \begin{split}
        = \sum_{nn'} \frac{(-1)^{n+n'}}{2l+1} \sum_{\nu} & a^{l}_{\nu,n} a^{l}_{\nu,-n'} \Big[A^{\Lambda*}_{ln} \Big]^{D} \Big[A^{\Lambda'}_{ln'} \Big]^{D} \\ & \times \mathrm{Re} \int_{-\infty}^{0} d\tau e^{-E^{l}_{\nu}\tau} e^{-im\omega_{0}|\tau|}.
        \end{split}
    \end{split}
\end{equation}
Performing the integral and inserting back into the relaxation superoperator, we have,
\begin{equation}
\begin{split}
    \hat{\Gamma}^{\Lambda \Lambda'}_{l} & = \delta_{ll'}\hat{\Gamma}^{\Lambda\Lambda'}_{ll'} \\
    & = -c^{\Lambda}c^{\Lambda'} \sum_{m} J^{\Lambda\Lambda'}_{lm}(\omega_{0}) \Big[\hat{X}^{\Lambda}_{lm}\Big]^{L} \Big[\hat{X}^{\Lambda'\dagger}_{lm}\Big]^{L},
\end{split}
\end{equation}
\noindent with,
\begin{equation}
\begin{split}
    J^{\Lambda \Lambda'}_{lm}(\omega_{0}) = \sum_{nn'}  & \frac{(-1)^{n+n'}}{2l+1} \Big[A^{\Lambda*}_{ln} \Big]^{D} \Big[A^{\Lambda'}_{ln'} \Big]^{D} \\
    & \times \sum_{\nu} a^{l}_{\nu,n} a^{l}_{\nu,-n'} \frac{E^{l}_{\nu}}{E^{(l)2}_{\nu}+m^{2}\omega^{2}_{0}}.
\end{split}
\end{equation}
%
%For a symmetric rotor, \(a^{l}_{\nu,n} = a^{l}_{\nu,-n'} = 1 \ \forall \ \nu = \pm n, \pm n'\), and the eigenvalues are \(E^{l}_{\nu} \xrightarrow{} E^{J}_{K}=D_{\perp}J(J+1) + (D_{\parallel}-D_{\perp})K^{2}\), where \(D_{\perp}\) and \(D_{\parallel}\) are eigenvalues of the diffusion tensor.
%
This is the general case of an asymmetric-top molecule. To derive the case for a symmetric rotor with the \(P\)- and \(D\)-frames coincident (as in our model), we refer back to eq. (\ref{diffusion_operator}), where we may write it in the \(D\)-frame as,

\begin{equation} \label{H_diff_op_symmetric}
    H_{\mathrm{rot-diff}} = D_{\perp} \left( L^{2}_{x} + L^{2}_{y} \right) + D_{\parallel} L^{2}_{z}
\end{equation}
%\begin{equation} 
   % H_{\mathrm{rot-diff}} = D_{\perp} \bigg\{ \frac{\partial^{2}}{\partial\beta^{2}} + \mathrm{cot}\beta\frac{\partial}{\partial}\beta + \bigg(\frac{D_{\parallel}}{D_{\perp}} + \mathrm{cot}^{2}\beta \bigg) \frac{\partial^{2}}{\partial \gamma^{2}}
   % & + \frac{1}{\mathrm{sin}^{2}\beta}\frac{\partial^{2}}{\partial \alpha^{2}} - \frac{2\mathrm{cos}\beta}{\mathrm{sin}^{2}\beta}\frac{\partial^{2}}{\partial \alpha \gamma} \bigg\},
%\end{equation}
%
where \(D_{\perp}\) and \(D_{\parallel}\) are rotational diffusion constants associated with axes perpendicular and parallel, respectively, with the spin chain.

Eq. (\ref{H_diff_op_symmetric}) is of the same form as the rigid-rotor Hamiltonian for a symmetric-top. As such, the eigenfunctions and eigenvalues in eq. (\ref{conditional_probability}) are those of a quantum mechanical rigid-rotor~\cite{zare_angular_1988,wang_anisotropic-rotational_1973}:
\begin{equation}
    \psi_{\nu} (t) \xrightarrow[]{} \phi^{J}_{K,M}(t) \equiv (-1)^{M-K}\sqrt{\frac{2J+1}{8\pi^{2}}} D^{(J)}_{-M-K}(\Omega(t))
\end{equation}
\begin{equation}
    E_{\nu} \xrightarrow[]{} E^{J}_{K} \equiv D_{\perp} J( J + 1 ) + K^{2} ( D_{\parallel} - D_{\perp} ).
\end{equation}
For our specific system, \(J \equiv l = 2\) and \(K = 0\). To see this, note that the only non-zero component for the dipole-dipole (DD) interaction is \( \left[ A^{\mathrm{DD}}_{20}\right]^{P} = \sqrt{6} \) in the \(P\)-frame. The \(z\)-principal axis is defined as a vector connecting the two nuclei in question. We then assume this is coincident with the \(z\)-axis in the \(D\)-frame, with the \(x\)- and \(y\)-axes arbitrary. Writing, \(D^{(l)}_{mn}(\alpha,\beta,\gamma) = e^{-im\alpha} d^{(l)}_{mn}(\beta) e^{-in\gamma} \) and assuming coincidence of the \(P\)- and \(D\)-frames (i.e. \(\{\alpha,\beta,\gamma\} = \{0,0,0\}\)), eq. (\ref{wigner_rotation_relation}) may be written,
\begin{equation}
\begin{split}
    \Big[ A^{\mathrm{DD}}_{20} \Big]^{P} & = \sum_{n} \Big[ A^{\mathrm{DD}}_{2n} \Big]^{D} d^{(2)}_{n0}(\beta = 0) \\
    & = \Big[ A^{\mathrm{DD}}_{20} \Big]^{D},
\end{split}
\end{equation}
\noindent where \(d^{(2)}_{00}(\beta) = \frac{3 \mathrm{cos}^{2}\beta - 1}{2}\) is the only non-vanishing reduced Wigner function, and equates to unity for \(\beta = 0\).

In general, the CSA \(P\)-frame will not be coincident with the DD \(P\)-frame. To simplify analytical expressions in the main text, however, they are assumed to be so for our system. Also, assuming cylindrical symmetry of the rigid-rod, the biaxiality parameters may be approximated as 0. Then the same relation holds for the CSA interaction; i.e. the only non-vanishing component of the spatial tensor is,
\begin{equation}
    \Big[ A^{\mathrm{CSA}}_{20} \Big]^{P} = \Big[ A^{\mathrm{CSA}}_{20} \Big]^{D}.
\end{equation}
Since \(n = K\), the only non-vanishing term in the correlation function is then \(E^{(2)}_{0} = 6 D_{\perp}\), and related to the rotational correlation time \(\tau_{\perp} \equiv \big(6D_{\perp}\big)^{-1}\). We then see that rotational motion about the principal axis of inertia parallel to the rod does not modulate the interactions responsible for relaxation, and this is a consequence of the coincidence of the \(P\)- and \(D\)-frames, as well as the assumption that the molecule tumbles rigidly. The secularised time-correlation function for our system becomes,
\begin{equation} \label{time_corr_function}
    G^{ll'}_{mm'nn'}(\tau) = \delta_{ll'}\delta_{mm'}\frac{(-1)^{n+n'}}{2l+1} e^{-\tau/\tau_{\perp}}.
\end{equation}

\begin{figure}[tbh] \label{relax_superop_plot}
\centering
\includegraphics[width=0.6\linewidth]{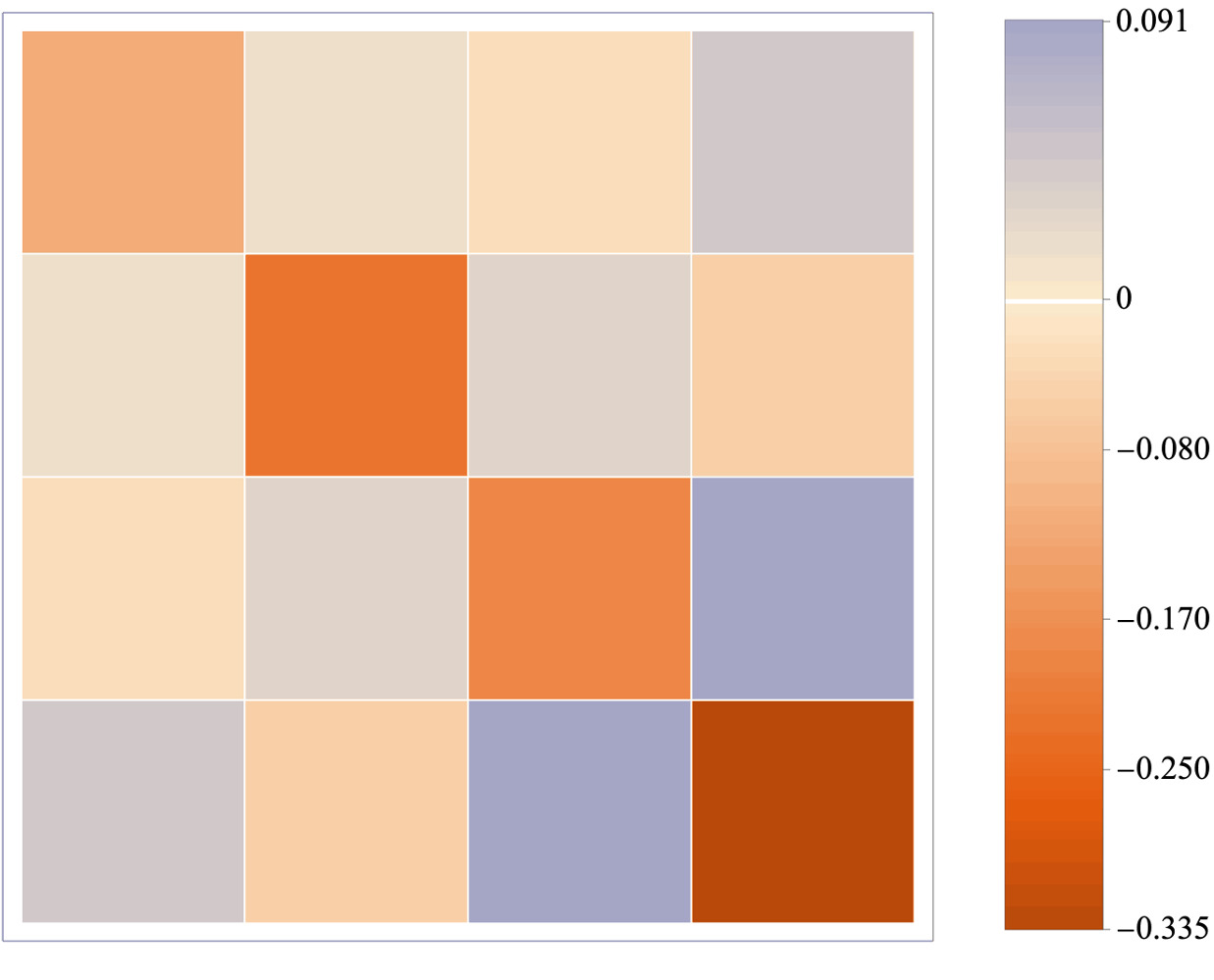}
\caption{ The \(\ (-1) \)-QC block of the real part of the Liouvillian in the \(\mathbb{B}_{Q}\) basis.
}
\label{fig:4x4_relax_superop_plot} 
\end{figure}

The \((-1)\)-QC subspace of the Liouvillian is given below. This shows that the off-diagonal elements are orders of magnitude smaller than the diagonal elements, and we, therefore, use the \(\mathbb{B}_{Q}\) basis and regard relaxation as a small perturbation when determining peak position. Also, since the mixing angle \(\theta = \mathrm{arctan}\left(\omega_{\Delta}/\omega_{J}\right) = -0.075\), the \(\mathbb{B}_{\mathrm{ST}}\) is used to simplify the \(\lambda_{\pm}\) expressions in the main text. The \((-1)\)-QC subspace of the Liouvillian is,
\begin{equation} \label{relax_supermatrix_4x4}
\hat{\mathcal{L}}_{4\times4} = 
\bordermatrix{ ~ & \big| |S'_{0} \rangle \langle T'_{+1} | \big) & \big| | T'_{-1} \rangle \langle S'_{0} | \big) & \big| | T'_{0} \rangle \langle T'_{+1} | \big) & \big| | T'_{-1} \rangle \langle T'_{0} | \big)  \cr
\big| |S'_{0} \rangle \langle T'_{+1} | \big) & -0.0852 + 1339.86 i & 13.43 \times 10^{-5} & -3.16 \times 10^{-3} & 3.47 \times 10^{-3} \cr
 \big| | T'_{-1} \rangle \langle S'_{0} | \big) &  13.43 \times 10^{-5} & -0.180 - 1339.86 i & 3.65 \times 10^{-3} &  -5.78 \times 10^{-3} \cr
 \big| | T'_{0} \rangle \langle T'_{+1} | \big) & -3.16 \times 10^{-3} & 3.65 \times 10^{-3} & -0.168 + 5.68 i &  94.16 \times 10^{-3} \cr
 \big| | T'_{-1} \rangle \langle T'_{0} | \big) & 3.47 \times 10^{-3} & -5.77 \times 10^{-3} & 94.16 \times 10^{-3} & - 0.338 - 5.68 i}
\ \mathrm{s}^{-1}.
\end{equation}
The real part is plotted is derived from the relaxation superoperator alone, and is plotted in figure \ref{fig:4x4_relax_superop_plot}.

The computed magnetic shielding tensors are given in their respective \(P\)-frames by:
\begin{equation} \label{mag_shielding_tensor_1}
\mathbf{\sigma}_{1} = 
\begin{pmatrix}
258.73 & 0 & 0 \\
0 & 41.59 & 0 \\
0 & 0 & 38.65 
\end{pmatrix}
\mathrm{ppm},
\end{equation}
\noindent and,
\begin{equation} \label{mag_shielding_tensor_2}
\mathbf{\sigma}_{2} = 
\begin{pmatrix}
258.64 & 0 & 0 \\
0 & 42.25 & 0 \\
0 & 0 & 38.90 
\end{pmatrix}
\mathrm{ppm}.
\end{equation}
These are transformed to the chemical shift tensor \(\delta\) by the relation,
\begin{equation}
    \mathbf{\delta} = \mathbb{I} \sigma_{\mathrm{TMS}}^{\mathrm{iso}} - \mathbb{\sigma},
\end{equation}
\noindent where \(\mathbb{I}\) is the three-dimensional identity matrix, and \(\sigma_{\mathrm{TMS}}^{\mathrm{iso}}\) is the isotropic component of the magnetic shielding tensor of tetramethylsilane, acting as a reference.

The shielding tensors (\ref{mag_shielding_tensor_1}) and (\ref{mag_shielding_tensor_2}) are transformed to their principal axis frames by diagonalisation. Then, the \textit{Haeberlen convention}~\cite{haeberlen_high_1976} is used to define the anisotropy and 
%asymmetry 
biaxiality parameters, respectively, as,
\begin{equation}
    \delta^{\mathrm{CSA}} = \delta^{\mathrm{P}}_{zz} - \delta^{\mathrm{iso}}
\end{equation}
and,
\begin{equation}
    \eta = \frac{\delta^{\mathrm{P}}_{xx}-\delta^{\mathrm{P}}_{yy}}{\delta^{\mathrm{CSA}}},
\end{equation}
with tensor components defined by,
\begin{equation}
    |\delta^{\mathrm{P}}_{zz} - \delta^{\mathrm{iso}}| \geq |\delta^{\mathrm{P}}_{xx} - \delta^{\mathrm{iso}}| \geq |\delta^{\mathrm{P}}_{yy} - \delta^{\mathrm{iso}}|.
\end{equation}

\subsection{Estimation of \(\tau_{\perp}\)}

The correlation time was estimated using the experimental \(T_{1} = 2.24 \ \mathrm{s}\) value and the relation,
\begin{equation}
    T^{-1}_{1} \simeq -\frac{\big( I_{z} \big| \hat{\Gamma} \big| I_{z} \big)}{\big( I_{z} \big| I_{z} \big)},
\end{equation}
and solving for \(\tau_{\perp}\).

\section{Synthesis of Target Triyne 5}

\begin{figure}[tbh]
\centering
\includegraphics[width=0.95\linewidth]{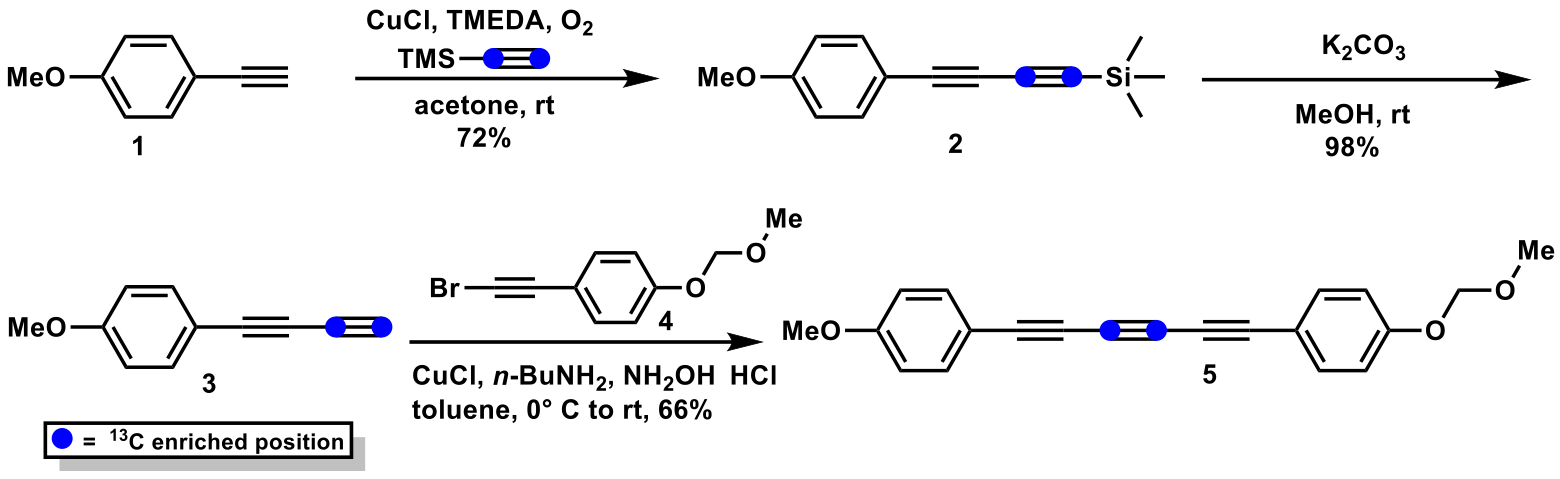}
\label{fig:synthesis_of_I} 
\end{figure}

\subsection{Synthesis of intermediate 2}

\begin{figure}[tbh]
\centering
\includegraphics[width=0.3\linewidth]{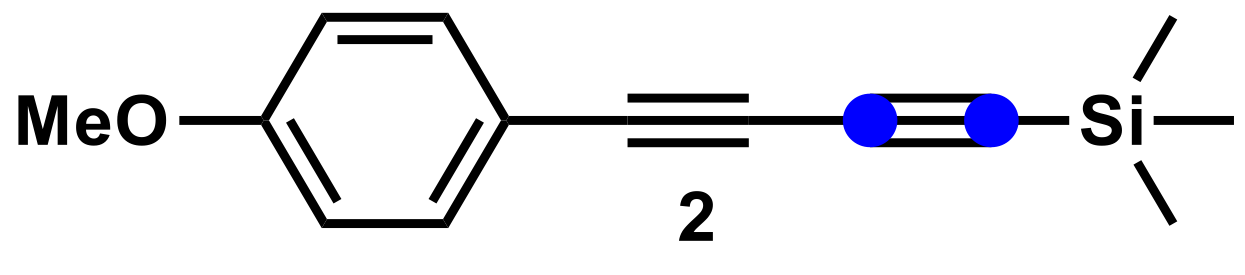}
\label{fig:intermediate_2} 
\end{figure}

To a stirred suspension of CuCl (41.6 mg, 0.42 mmol) in acetone (4 mL) was added tetramethylethylenediamine (TMEDA, 22.0 \textmu L, 0.14 mmol), and the mixture was stirred for 30 minutes. Then, a mixture of 1-ethynyl-4-methoxybenzene \textbf{1} (264.3 mg, 2.0 mmol) and (trimethylsilyl)acetylene-\(^{13}\mathrm{C}_{2}\) (288 \textmu L, 2.0 mmol) in acetone (4 mL) was added slowly and bubbled with \(\mathrm{O}_{2}\) for 5 min. The reaction mixture was stirred for 2 h at room temperature, and then passed through a pad of silica gel. Then, the filtrate was concentrated under reduced pressure and the residue was purified by column chromatography using \(\mathrm{Et}_{2}\mathrm{O/hexane}\) (1:15) as eluent to afford compound \textbf{2} (332 mg, 72\%) as a colorless oil. \(^{1}\mathrm{H}\) NMR (400 MHz, \(\mathrm{CDCl}_{3}\)): \(\delta\) 0.24 (dd, \(J=2.5\), 0.6 Hz, 9 H), 3.82 (s, 3 H), 6.84 (d, \(J=9.4\) Hz, 2 H), 7.44 (d, \(J=9.4\) Hz, 2 H). \(^{13}\mathrm{C}\{^{1}\mathrm{H}\}\) NMR (101 MHz, \(\mathrm{CDCl}_{3}\)): \(\delta\) 87.8 (d, \(J=144.5\) Hz, 1 C), 90.1 (d, \(J=144.5\) Hz, 1 C) (only \(^{13}\mathrm{C}\)-enriched signals are shown). LRMS (ES+) \(m/z\) 231.1 (100\%, [M + H]\(^{+}\)).

\subsection{Synthesis of intermediate 3}

\begin{figure}[tbh]
\centering
\includegraphics[width=0.3\linewidth]{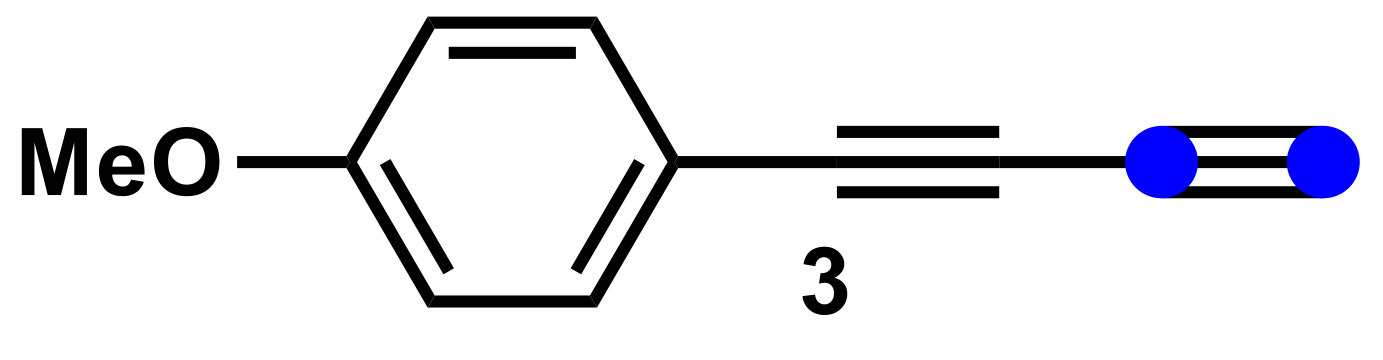}
\label{fig:intermediate_3} 
\end{figure}

A mixture of \textbf{2} (320 mg, 1.39 mmol), \(\mathrm{K}_{2}\mathrm{CO}_{3}\) (384 mg, 2.78 mmol), MeOH (5 mL) and THF (5 mL) was stirred at room temperature for 1 h. Then, the reaction mixture was extracted with ethyl acetate and washed with brine. Evaporation of the  solvent afforded diacetylene \textbf{3} (215 mg, 98\%) as a colorless solid, which was used for the next step without further purification. \(^{1}\mathrm{H}\) NMR (400 MHz, \(\mathrm{CDCl}_{3}\)): \(\delta\) 2.46 (dd, \(J=232.8\), 77.8 Hz, 1H), 3.83 (s, 3H), 6.85 (d, \(J=8.9\) Hz, 1 H), 7.47 (d, \(J=8.93\) Hz, 2 H). \(^{13}\mathrm{C}\{^{1}\mathrm{H}\}\) NMR (101 MHz, \(\mathrm{CDCl}_{3}\)): \(\delta\) 68.03 (d, \(J=192.20\) Hz, 1 C), 71.04 (d, \(J=191.47\) Hz, 1 C) (only \(^{13}\mathrm{C}\)-enriched signals are shown). LRMS (ES+) \(m/z\) 159.1 (100\%, [M + H]\(^{+}\)).

\subsection{Synthesis of triyne I}

\begin{figure}[tbh]
\centering
\includegraphics[width=0.5\linewidth]{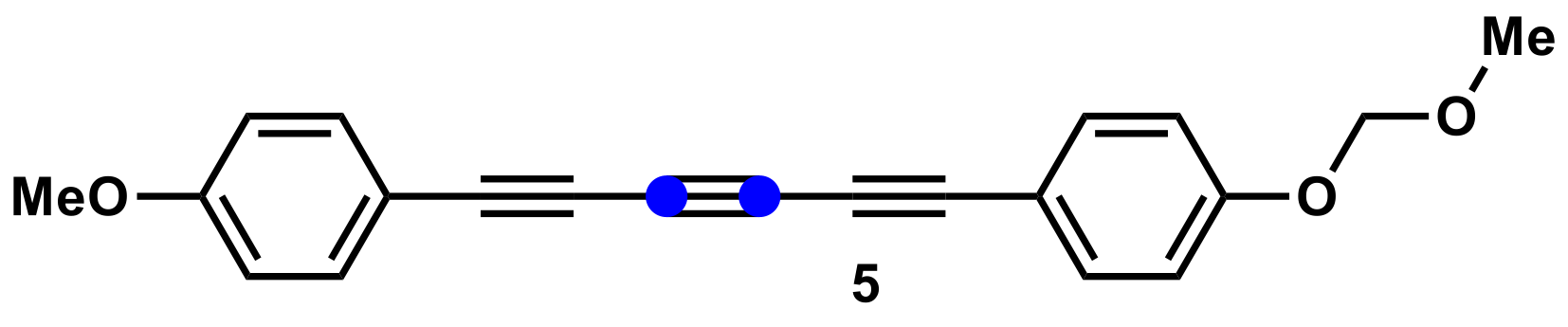}
\label{fig:triyne_5} 
\end{figure}

To a solution of diacetylene \textbf{3} (197 mg, 1.26 mmol) in toluene (2.5 mL) at 0 °C were added CuCl (18.8 mg, 0.19 mmol), \(\mathrm{NH}_{2}\mathrm{OH}\bullet\mathrm{HCl}\) (26.3 mg, 0.38 mmol) and \(n\)-\(\mathrm{BuNH}_{2}\) (188 \textmu L, 1.9 mmol) in order. Alkynyl bromide \textbf{4} (304 mg, 1.26 mmol) was diluted with 2.5 mL toluene and was added dropwise to the mixture. The reaction mixture was allowed to warm to room temperature and stir for 18 h. The reaction was quenched with \(\mathrm{H}_{2}\mathrm{O}\) and extracted with ether. The organic layer was washed with \(\mathrm{H}_{2}\mathrm{O}\), brine and dried over \(\mathrm{MgSO}_{4}\). The solvent was evaporated, and the residue was purified by column chromatography using dichloromethane/hexane (1:2) as eluent to afford triyne \textbf{I} (263 mg, 66\%) as a colorless solid. \(^{1}\mathrm{H}\) NMR (400 MHz, \(\mathrm{CDCl}_{3}\)): \(\delta\) 3.48 (s, 3 H), 3.84 (s, 3 H), 5.20 (s, 2 H), 6.86 (d, \(J=8.0\) Hz, 2 H) 7.00 (d, \(J=8.0\) Hz, 2 H), 7.50 (d, \(J=8.9\) Hz, 2 H), 7.49 (d, \(J=8.9\) Hz, 2 H). \(^{13}\mathrm{C}\{^{1}\mathrm{H}\}\) NMR (101 MHz, \(\mathrm{CDCl}_{3}\)): \(\delta\) 66.41 (d, \(J=217.0\) Hz, 1 C), 66.39 (d, \(J=217.0\) Hz, 1 C) (only \(^{13}\mathrm{C}\)-enriched signals are shown). LRMS (ES+) \(m/z\) 317.1 (100\%, [M + H]\(^{+}\)).

\bibliographystyle{rsc}
\bibliography{references/references.bib}

%% file: symbols/main.tex
\input{symbols/notes}
\input{symbols/chem}
\input{symbols/interactions}
\input{symbols/Euler}
\input{symbols/bases}

%% file: symbols/notes.tex
%Coloured comments inside the main document.
%
\newcommand{\blue}[1]{\color{blue}{#1}\color{black}\xspace}
\newcommand{\red}[1]{\color{red}{#1}\color{black}\xspace}
\newcommand{\purple}[1]{\color{purple}{#1}\color{black}\xspace
}
\newcommand{\green}[1]{\color{olive}{#1}\color{black}\xspace
}
%______________________________________________
\newcommand\MHLnote[1]{\blue{[MHL: #1]}\xspace}
\newcommand\CBnote[1]{\red{[CB: #1]}\xspace}
\newcommand\JWnote[1]{\green{[JW: #1]}\xspace}
%______________________________________________
\newcommand{\ToHere}{\ \newline \MHLnote{** To Here **} \newline}

%% file: symbols/chem.tex
\newcommand{\Cth}{${}^{\mathrm{13}}\mathrm{C}$\xspace}
\newcommand{\Ctwo}{${}^{\mathrm{13}}\mathrm{C}_\mathrm{2}$\xspace}
\newcommand{\One}{\textbf{I}\xspace}
\newcommand{\CDCl}[1]{$\mathrm{CDCl_{#1}}$\xspace}

%% file: symbols/interactions.tex
\newcommand{\JCC}{J_\mathrm{CC}}

%% file: symbols/Euler.tex
\newcommand{\WDL}{\Omega_{DL}}
\newcommand{\aDL}{\alpha_{DL}}
\newcommand{\bDL}{\beta_{DL}}
\newcommand{\gDL}{\gamma_{DL}}
\newcommand{\WPjkD}{\Omega_{PD}^{jk}}
\newcommand{\aPjkD}{\alpha_{PD}^{jk}}
\newcommand{\bPjkD}{\beta_{PD}^{jk}}
\newcommand{\gPjkD}{\gamma_{PD}^{jk}}
\newcommand{\WPjD}{\Omega_{PD}^{j}}
\newcommand{\aPjD}{\alpha_{PD}^{j}}
\newcommand{\bPjD}{\beta_{PD}^{j}}
\newcommand{\gPjD}{\gamma_{PD}^{j}}
\newcommand{\WPkD}{\Omega_{PD}^{k}}
\newcommand{\aPkD}{\alpha_{PD}^{k}}
\newcommand{\bPkD}{\beta_{PD}^{k}}
\newcommand{\gPkD}{\gamma_{PD}^{k}}

%% file: symbols/bases.tex
\newcommand{\BST}{\mathbb{B}_{\mathrm{ST}}}
\newcommand{\BSTp}{\mathbb{B}'_{\mathrm{ST}}}

%% file: figures/triyne_and_tensors.tex
\begin{figure*}
\centering
\includegraphics[width=0.9\linewidth]{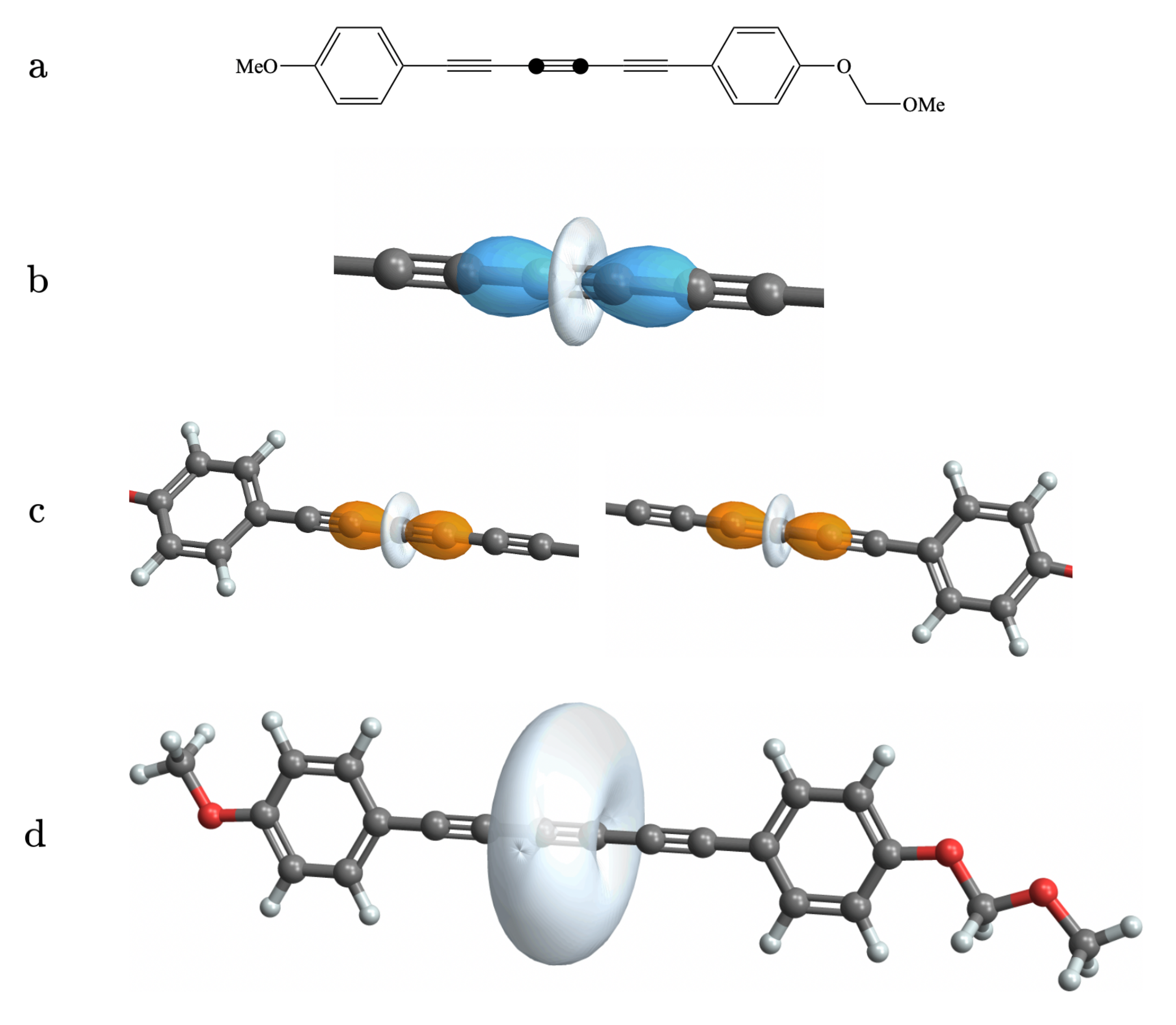}
\caption{
(a) Molecular structure of \One, with \Cth labelled sites depicted by black circles; (b) The rank-2 part of the \Cth-\Cth dipole-dipole coupling tensor,
%associated with the labelled carbons 
represented by an ovaloid~\cite{radeglia_ovaloid_1995,young_tensorview_2019}; (c) The calculated \Cth CSA tensors of the \Cth labels represented by ovaloids; (d) The inertia tensor of the molecule, represented as an ovaloid,  superimposed on the molecular structure. The grey atoms are C, the red atoms O, and white H. The graphics were generated in \emph{SpinDynamica}~\cite{bengs_spindynamica_2018}.
}
\label{fig:triyne_and_tensors} 
\end{figure*}

%% file: figures/full_and_superimposed_spectrum.tex
\begin{figure}[tbh]
\centering
\includegraphics[width=0.9\linewidth]{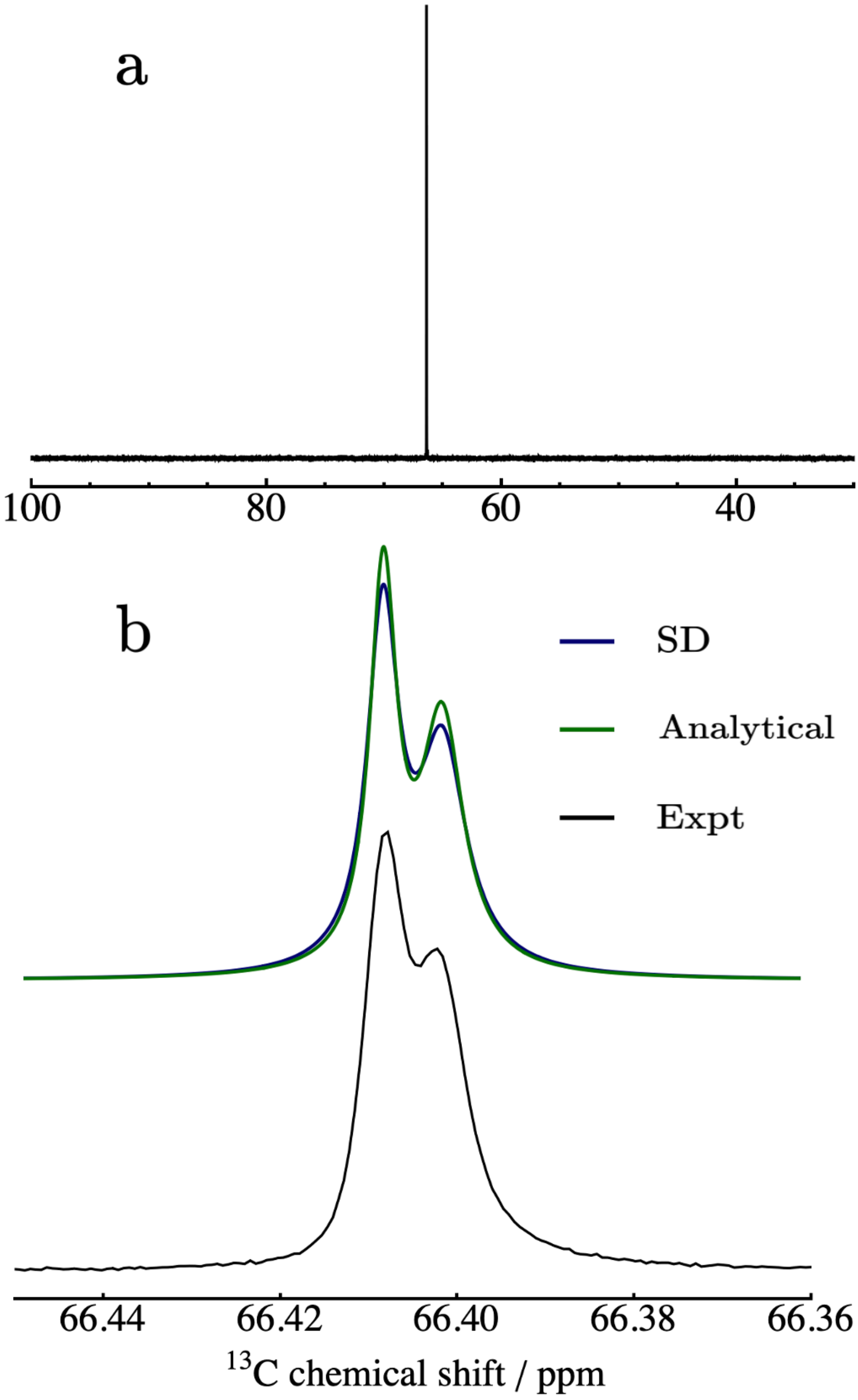}
\caption{\Cth spectra of a 0.3~M solution of \One in \CDCl3, at a magnetic field of 9.4 T. (a) Overview of the \Cth spectrum; (b) Black line: Expanded view of the central doublet, showing the strongly asymmetric linewidths of the doublet components. Dark blue line: Numerical \emph{SpinDynamica} simulation~\cite{bengs_spindynamica_2018}, using the theory given in the text and parameters in table \ref{tab:parameters}. Green line:
Superposition of two Lorentzians with amplitudes, frequencies, and linewidths specified by table \ref{tab:analytical_results} and eq. (\ref{approximate_rates}).
%\MHLnote{The following phrase (in blue) is too obscure. Specify \emph{exactly} which equations are used.}
%\blue{using approximations leading to eq. (\ref{approximate_rates}), with the exception of the extreme narrowing limit.}
%\MHLnote{I'm assuming that the analytical expressions, when corrected, match very closely the numerical simulation and experimental data.  }
%\MHLnote{do not use an equation number in the graphics. This is bound to lead to trouble.}
%
}
\label{fig:full_and_superimposed_spectrum} 
\end{figure}

%% file: tables/parameters.tex
\begin{table}[bt]
\caption{\label{tab:parameters}
Spin system parameters for \One in solution.}
\begin{ruledtabular}
\begin{tabular}{lll}
Parameter & Value & Note  
\\[5pt]\hline
\\[2pt]
$J_{jk}$ & 214.15 Hz & \footnotesize{Experimental $^a$}
\\[5pt]
$\Delta\delta_\mathrm{iso}$ & 0.16 ppm
& \footnotesize{Experimental $^b$}
\\[5pt]
$b_{jk}/(2\pi)$ & \(-\)4152.84 Hz & \footnotesize{Estimated $^c$}
\\[5pt]
$\{\aPjkD,\bPjkD,\gPjkD\}$ & $\{0,-2.5,0\}^\circ$
& \footnotesize{Frames obtained by diagonalising} \\
& & \footnotesize{calculated tensors}
\\[5pt]
$\delta_j^\mathrm{CSA}$ & $-145.7$ ppm
& \footnotesize{Calculated}
\\[5pt]
$\eta_{j}$ & 0.020 & \footnotesize{Calculated}
\\[5pt]
$\{\aPjD,\bPjD,\gPjD\}$ & $\{0,-2.6,0\}^\circ$
& \footnotesize{Frames obtained by diagonalising} \\
& & \footnotesize{calculated tensors}
\\[5pt]
$\delta_k^\mathrm{CSA}$ & $-145.4$ ppm
& \footnotesize{Calculated}
\\[5pt]
$\eta_{k}$ & 0.023 & \footnotesize{Calculated}
\\[5pt]
$\{\aPkD,\bPkD,\gPkD\}$ & $\{0,-2.6,0\}^\circ$
& \footnotesize{Frames obtained by diagonalising} \\
& & \footnotesize{calculated tensors}
\\[5pt]
$\tau_\perp$ & \(136.5\) ps
& \footnotesize{Estimated from the parameters in this} \\  
& & \footnotesize{table and experimental $T_{1}$
}
\\[2pt]
\end{tabular}
\end{ruledtabular}
\footnotesize{$^a$ Obtained from \(90^\circ\) pulse-acquire spectrum on a 700 MHz spectrometer. $^b$ Estimated from the \Cth spectrum of natural abundance material; $^c$ Estimated from the internuclear distance, $r_{jk}=122\,\  \mathrm{pm}$.
}\\
\end{table}

%% file: figures/frame_transformations.tex
\begin{figure*}
\centering
\includegraphics[width=0.9\linewidth]{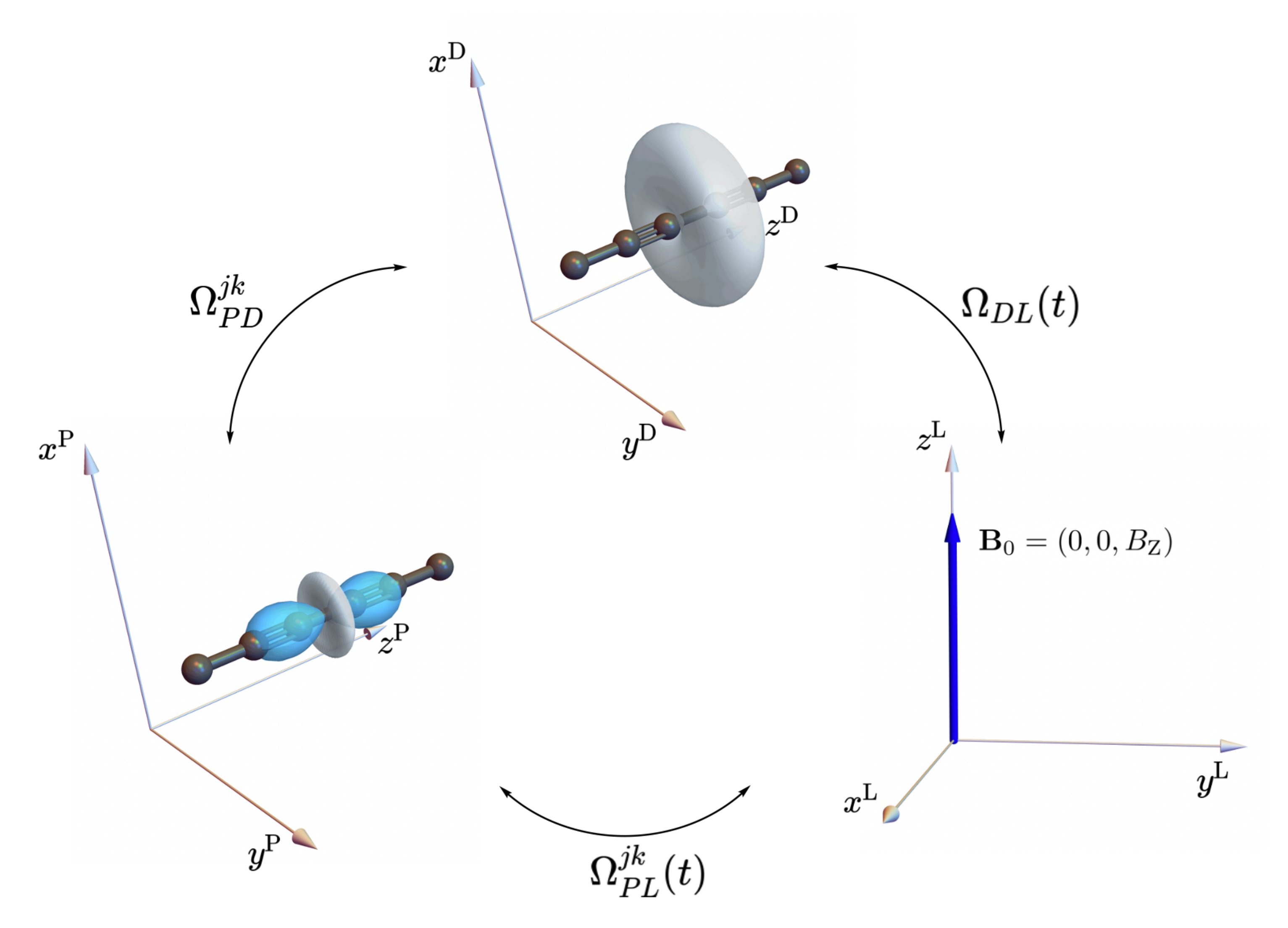}
\caption{
Relevant frame transformations illustrated using the DD-tensor as an example. On the left, the coordinate system is the molecule-fixed \(P\)-frame of the DD interaction, with the \(z\)-axis parallel to the internuclear vector. The set of angles $\WPjkD$ orientate the \(P\)- and \(D\)-frames. The molecule-fixed \(D\)-frame is given  by the principal axis frame of the inertia tensor with its \textit{z}-axis parallel to the molecular long axis. The laboratory frame L is defined such that its \textit{z}-axis is parallel to the applied magnetic field. 
The angles \(\Omega_{DL}(t)\) orient the \(D\)- and \(L\)-frames with respect to each other. These angles are time-dependent, since the \(L\)-frame is space-fixed and stochastic molecular tumbling continuously alters the orientation of the \(D\)- and \(L\)-frames with respect to one another. The angles parameterising the transformation between the \(P\)- and \(L\)-frames will be time-dependent for the same reason.
}
\label{fig:frame_transformations} 
\end{figure*}

%% file: tables/tensor_components.tex
\begin{table}[bt]
\caption{\label{tab:tensor_components}
Irreducible spherical spatial tensor components for \(l=2\), \(p=0\) in their principal axis frame~\cite{smith_hamiltonians_1992-1}}
\begin{ruledtabular}
\begin{tabular}{lccc}
Interaction, \(\Lambda\)   &\(c^{\Lambda}\)   &\(\big[ A^{\Lambda}_{20}\big]^{P}\) \\[5pt]
\hline
\\
DD, spins $I_j$ and $I_k$       & \(b_{jk}\)            &\(\sqrt{6}\)    \\
\\
CSA, spin $I_j$     &\(-\gamma_{j}\)     &\(\sqrt{\frac{3}{2}}\delta_j^\mathrm{CSA}\)  \\
\\
\end{tabular}
\end{ruledtabular}
\end{table}

%% file: tables/spin_tensor_components.tex
\begin{table}[bt]
\caption{\label{tab:spin_tensor_components}
Irreducible spherical spin and spin-field tensor components for \(l=2\) in the L-frame~\cite{smith_hamiltonians_1992-1}}
\begin{ruledtabular}
\begin{tabular}{lccc}
Interaction, \(\Lambda\)   &\(m\)    &\(\big[ X^{\Lambda}_{2m} \big]^{L}\) \\[5 pt]
\hline
\\
\multirow{6}{4em}{DD, spins $I_j$ and $I_k$}  & 0   & \( \frac{1}{2\sqrt{6}} (4 I_{jz} I_{kz} - I^{-}_{j} I^{+}_{k} - I^{+}_{j} I^{-}_{k}) \)   \\
                \\
                 & \(\pm 1\)   & \(\mp \frac{1}{2}( I^{\pm}_{j} I_{kz} + I_{jz} I^{\pm}_{k} )\) \\
                \\
                 & \(\pm 2\)   & \( \frac{1}{2}(I^{\pm}_{j} I^{\pm}_{k})\) \\
                 \\
\multirow{6}{4em}{CSA, spin $I_j$}  & \(0\)  & \(\sqrt{\frac{2}{3}} B_{0}I_{jz}\)  \\
\\
                        & \(\pm 1\) & \(\mp \frac{1}{2} B_{0} I^{\pm}_{j}\) \\
                        \\
                        & \(\pm 2\) & \(0\) \\
                        \\
\end{tabular}
\end{ruledtabular}
\end{table}

%% file: tables/eigenoperators.tex
\begin{table}[bt]
\caption{\label{tab:coherence_eigen_op_vals}
Coherence eigenoperators of \(\hat{H}_{\mathrm{coh}}\) along with the associated eigenfrequencies and peak amplitudes.
%\MHLnote{I've arranged the frequencies in a different way, to make it easier to see what they mean. For example the sign of $\omega_{\Sigma}$ is maintained. Can you check that I did not make any mistakes?}
%\JWnote{checked.}
}
\begin{ruledtabular}
\begin{tabular}{lcc}
\(\big|Q_{q}\big)\)   &\(\omega_{q}\)  &\(a_{q}\) \\
\hline
\\
\(\big| | S_{0}' \rangle \langle T_{+1}' |\big)\) & \(\frac{1}{2} 
\big( 
\omega_{\Sigma} + \omega_{J} + \omega_{e}  \big)
\)           &
\(\frac{1}{2}\mathrm{sin}^{2}\frac{\theta}{2}\) \\
\\
\(\big| | T_{-1}' \rangle \langle S_{0}' |\big)\)    &\(\frac{1}{2} \big( 
\omega_{\Sigma} -\omega_{J} -\omega_{e} 
\big)\)     &\(\frac{1}{2}\mathrm{sin}^{2}\frac{\theta}{2}\) \\
\\
\(\big| | T_{0}' \rangle \langle T_{+1}' |\big)\) &\( \frac{1}{2} 
\big( \omega_{\Sigma} + \omega_{J} - \omega_{e}  \big)\)    
&\(\frac{1}{2}\mathrm{cos}^{2}\frac{\theta}{2}\) \\
\\
\(\big| | T_{-1}' \rangle \langle T_{0}' |\big)\) &\( \frac{1}{2} \big(
\omega_{\Sigma} - \omega_{J} + \omega_{e} 
\big)\) 
&\(\frac{1}{2}\mathrm{cos}^{2}\frac{\theta}{2}\) \\
\\
\end{tabular}
\end{ruledtabular}
\end{table}

%% file: figures/coherences_with_spectrum.tex
\begin{figure}[tb]
\centering
\includegraphics[width=1.0\linewidth]{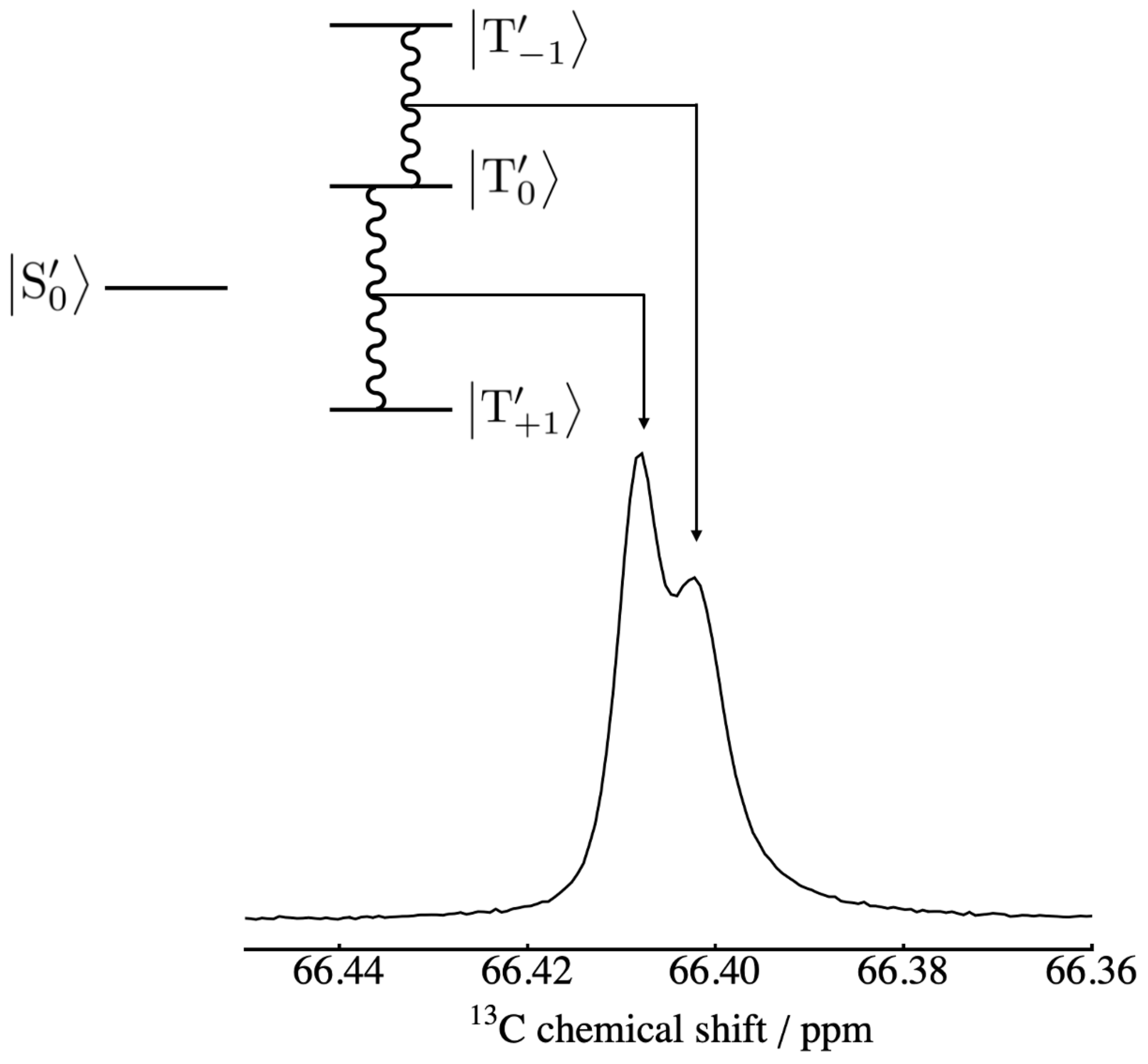}
\caption{The correspondence between the single-quantum triplet-triplet coherences (wiggly lines) and the NMR spectrum. 
%The analysis in section \ref{liouvillian_section} allowed the deduction that 
The coherence represented by the operator \(Q_{+}\) is associated with the narrow peak while the coherence represented by \(Q_{-}\) is associated with the broad peak.
%Note that coherences exist between the singlet- and outer triplet-states, but these are vanishingly small and have been omitted for simplicity.
}
\label{fig:coherences_with_spectrum} 
\end{figure}

%% file: figures/lambda_vs_B0_plots.tex
\begin{figure}[tb] \label{relax_superop_plot}
\centering
\includegraphics[width=0.9\linewidth]{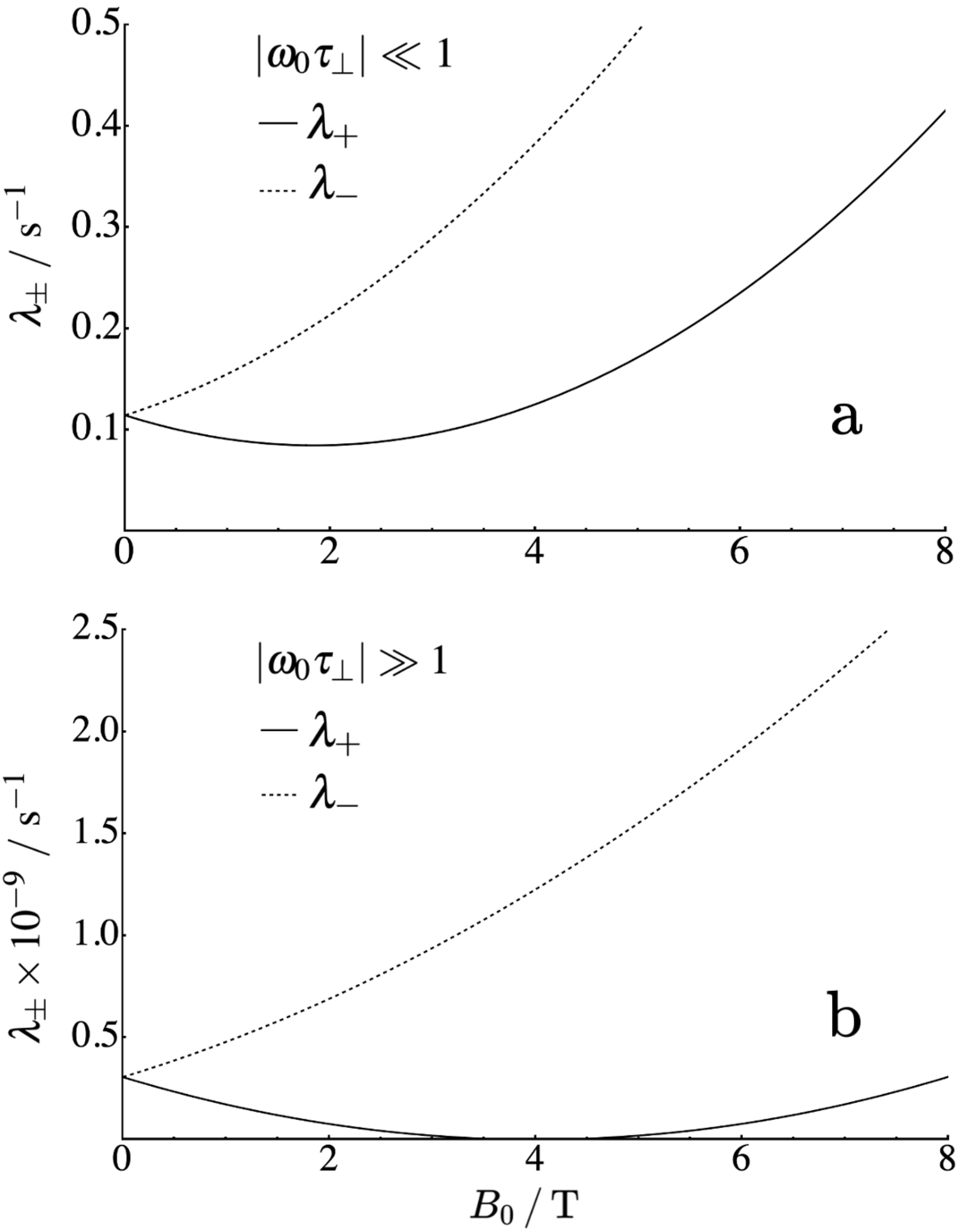}
\caption{
Plots of the linewidth parameters \(\lambda_{\pm}\) against external static field, for the parameters in table~\ref{tab:parameters}. (a) The extreme-narrowing limit, based on eq.~(\ref{approximate_rates}), showing the minimum \(\lambda_{+} = 8.47 \times 10^{-2} \ \mathrm{s}^{-1}\) at \(B_{0} = 1.84 \ \mathrm{T}\). (b) The long-\(\tau_{\perp}\) limit, with a minimum \(\lambda_{+} = 0 %\mathrm{s}^{-1}
\) 
at \(B_{0} = 4.0 \ \mathrm{T}\). The DD and CSA mechanisms cancel in the long-\(\tau_{\perp}\) limit at this magnetic field. The cancellation is incomplete in the extreme-narrowing limit.
%This doesn't happen in the extreme narrowing limit, however.
}
\label{fig:lambda_vs_B0_plots} 
\end{figure}

%% file: tables/analytical_results.tex
\begin{table}[bt]
\caption{\label{tab:analytical_results}
Parameters used to plot the analytical spectral function in fig. \ref{fig:full_and_superimposed_spectrum}.
}
\begin{ruledtabular}
\begin{tabular}{lll}
Parameter & Value & Note  
\\[5pt]\hline
\\[2pt]
\(\lambda_{+}\) & \( 0.583 \ \mathrm{ \ s^{-1}}\) & eq. (\ref{approximate_rates})
\\[5pt]
\(\lambda_{-}\) & \( 1.19 \ \mathrm{ \ s^{-1}}\) & eq. (\ref{approximate_rates})
\\[5pt]
\(a_{\pm}\) & \( 0.499\) & eq. (\ref{peak_amplitude})
\\[5pt]
\(\omega_{\pm}\) & \(\mp 1.90 \ \mathrm{rad \ s^{-1}}\) & eq. (\ref{eq:frequencies})
\\[5pt]
\end{tabular}
\end{ruledtabular}
\end{table}